\documentclass{aa}
\usepackage[varg]{txfonts}
\usepackage{hyperref}

\newcommand{\real}{\mathbb R}
\newcommand{\intl}{\int\limits}
\DeclareMathOperator{\sgn}{sgn}
\DeclareMathOperator{\ra}{ra}
\DeclareMathOperator{\dec}{dec}
\DeclareMathOperator{\data}{data}
\DeclareMathOperator{\diag}{diag}
\DeclareMathOperator*{\argmax}{argmax}

\begin{document}

\title{Rotation dynamics and torque efficiency of cometary nuclei}
\titlerunning{Rotation dynamics and torque efficiency}
\author{Matthias L\"auter\inst{\ref{inst1}} \and Tobias Kramer\inst{\ref{inst2}}}

\institute{%
Zuse Institute Berlin, 14195 Berlin, Germany, \email{laeuter@zib.de}\label{inst1}
\and
Institute for Theoretical Physics, Johannes Kepler Universit\"at Linz, Austria\label{inst2}
}
\date{Received: 21 January 2025 / Accepted: 30 May 2025 / Published: 04 July 2025 }

\dedication{\normalfont\small Published in Astronomy \& Astrophysics, Volume 699, A75,
  \href{https://doi.org/10.1051/0004-6361/202553845}{doi:10.1051/0004-6361/202553845}}

\abstract
{
The dynamics of a rigid cometary nucleus is described by the evolutions of its center-of-mass and of its rotation state.
Solar irradiation that reaches the surface of a cometary nucleus causes the sublimation of volatiles that form the coma around the nucleus.
The sublimation process transfers linear momentum and rotational angular momentum from the nucleus to the surrounding space, and thus affects the dynamics via nongravitational forces and nongravitational torques.
With the exception of close approaches to planets, these torques exert the dominant influence on the rotation states of cometary nuclei.
}
{
The 2014-2016 Rosetta mission accompanying the comet 67P/Churyumov-Gerasimenko provides the longest continuous observational data to track its rotation state.
In particular, the data set encompasses the direction of the angular velocity, denoted by $\omega$, and the angular frequency, $|\omega|$, over a time period of approximately 700 days.
The observed change in the rotation state is not explained by a low heat conductivity thermophysical model in combination with a homogeneous surface ice coverage of comet 67P.
Spatially and/or temporally varying weights for effective active fraction with respect to a prescribed set of surface regions provide a potential solution to this problem.
}
{
Here, we present a methodology for classifying the surface based on vectorial efficiency of the torque.
On any cometary surface without geometric symmetry, the methodology highlights the decomposition into eight characteristic regions that encode the signs of torque efficiency with respect to all vector components.
This decomposition is divided into two subsets of four regions, each of which is located in one of both hemispheric regions.
}
{
We analyze in detail rotation states close to lowest energy and different thermophysical models, and we discuss how the uncertainties of observations affect the model parameters.
We study the occurrence of these regions for an oblate ellipsoid, a near-prolate ellipsoid, a bilobed shape, and a shape model analogous to that of comet 67P.
The sensitivity analysis for comet 67P indicates that the observations constrain only one of the eight weights uniquely.
The other directions are poorly constrained and show the limitation of the rotational data in determining the regional activity on comet 67P.
}
{}


\maketitle

\nolinenumbers

\section{Introduction}

The sublimation of ice from a cometary nucleus transfers momentum from the nucleus to the surrounding space.
The spatial integration over the momentum field on the surface results in nongravitational acceleration (NGA) and nongravitational torque (NGT) acting on the rigid body of the comet.
The NGA leads to variations in osculating elements and in trajectories through the solar system, while the NGT changes the rotational angular momentum, and thus the rotation state of the nucleus.
The overall magnitudes of the NGA and the NGT depend on the effective cross section of the nucleus illuminated by the sun.

The significance of NGAs and in particular of NGTs for the change in the rotation state is pointed out by \cite{Whipple1950}.
Nongravitational torques are investigated; for example, for comet 2P/Encke by \cite{Whipple1979}, for comet 22P/Kopff by \cite{Sekanina1984}, for comet 1P/Halley by \cite{Peale1989}, \cite{Julian1990}, and for Halley-like comets by \cite{Samarasinha1995}.
For rotational dynamics of general comets, reviews are outlined by \cite{Samarasinha2004}, \cite{Yeomans2004}, and \cite{Knight2024}.

A rotating jet model on a spherical body used by \cite{Chesley2004} features NGAs but NGTs are always zero.
The torque formation and changes in rotation state vary with the underlying geometry; for example, oblate or prolate ellipsoids lead to different torque effects.
\cite{Sitarski1995} controls the evolution of the spin axis with an additional parameter in a linear precession model.
\cite{Whipple1979} and subsequent authors, for example \cite{Sekanina1981}, \cite{Krolikowska1998}, and \cite{Krolikowska2001}, study the evolution of the rotation axis for a nonspherical nucleus under the assumption of a forced precession model.
\cite{Jorda2002} and \cite{Gutierrez2002} investigate the sublimation activity of irregularly shaped cometary nuclei.
\cite{Kramer2019a} develop a Fourier series-based approach to parametrize the observed torque variations of comet 67P/Churyumov-Gerasimenko (67P/C-G) in a body-fixed coordinate system and compared the observations with different sublimation models.
\citet{Attree2019,Attree2023a,Attree2024} discuss the evolution of the torque of comet 67P/C-G based on the geomorphological properties defined in \cite{Marschall2017}. 
\cite{Davidsson2004} approximate the surface of an ellipsoid with a triangular mesh and regionally weighted activity.
\cite{Maquet2012} weight the sublimation activity based on latitudinal bands on an ellipsoid surface.

The rotational dynamics of a cometary nucleus can be considered as the rotation of a rigid body, as is pointed out by \cite{Samarasinha2004}.
Early approaches such as that of \cite{Whipple1979} are restricted to simple rotations, which are principal axis rotations.
These are characterized as rotations around only one of the principal axes of the tensor of inertia.
In this situation the directions for the rotational angular momentum and for the angular velocity are the same as that of the principal axis.
A special case of a principal axis rotation is the state of lowest rotational energy for a prescribed angular momentum.
It is the rotation around the axis of the largest moment of inertia.
In general, a nucleus can be in a nonprincipal axis rotation where the vectors for rotational angular momentum and for angular velocity do not point into the same direction.
As is discussed by \cite{Julian1987}, \cite{Samarasinha1995}, \cite{Gutierrez2003}, and \cite{Knight2024}, nonprincipal axis rotations can be in short axis mode or in long axis mode.
The lowest-energy state is the limiting case of a short axis mode.
As is described by \cite{Jewitt1997} and \cite{Frouard2018}, in the presence of any damping for kinetic energy, a cometary nucleus in an excited state (i.e., one that is above lowest energy) will always tend to reach a  lowest-energy state.
The two comets 1P and 103P are regarded as undergoing nonprincipal axis rotation, with the possibility that other comets may also exhibit this behavior, as is proposed by \cite{Knight2024}. 
Conversely, the majority of comets do not exhibit indications of strongly excited rotation states.
In the case of comet 67P/C-G, \cite{Gutierrez2016}, \cite{Godard2017}, and \cite{Kramer2019a} report only a minor excitation from the lowest-energy state.
This and the fact that most comets do not show excited rotations \citep{Knight2024} leads us to the assumption that the lowest-energy state reflects the properties of most known comets in an appropriate manner.
In this case, observational data for angular velocity can be used to assess torque directions.

Momentum is created by sublimation processes close to the surface of the cometary nucleus.
Incoming solar irradiation heats up the upper surface layers and is mostly reradiated into space.
The sublimation process consumes some fraction of the incoming energy, while the remaining energy propagates into subsurface layers, where other frozen volatiles might be present.
Thermophysical models (TPMs) describe these processes at a point on the surface of the nucleus, as is discussed by \cite{Keller2015a} and \cite{Blum2018}.
Complex TPMs consider the composition of the material on the surface and within the column of the subsurface domain, the temporal evolution of solar irradiation, and the energy fluxes within the column.
The analyses of the Rosetta images of comet 67P/C-G by \cite{Kramer2016a}, \cite{Gundlach2020}, and \cite{Fulle2020} point to very low values of heat conductance, and thus sublimation of water occurs almost instantaneously with respect to the incoming radiation.
Due to a higher sublimation rate at lower temperatures, CO${}_2$ emission can occur even at night.
Here, we constrict the discussion to TPMs that establish sublimation flux as an instantaneous response to the incoming irradiation.
These models balance reradiation due to scattering or thermal activity and sublimation energy.
Heat transfer into the cometary material is neglected.
An increase in sublimation is related to an increase in irradiation (monotonous function).
In this case, no heat flow to the subsurface layers or any other delaying process is considered.
An example of such an approach is model A in \cite{Keller2015}, which is based on the energy balance for incoming radiation and outgoing thermal and sublimation fluxes on the surface.

The shape of the nucleus is of particular importance for rotational dynamics.
In addition to the tensor of inertia, the formation of torques is strongly affected by the local geometry of the surface faces with respect to the rotational axis.
The basic geometry-driven effects can already be studied for simplified ellipsoidal shapes.
More complex body shapes further complicate the analysis, for instance, by introducing concavities.
Concavities change the irradiation by self-shadowing and lead to significant variability in the regional torque directions.
The basic effect of concavities can be studied for a shape with a simplified bilobed structure.
Shapes that approximate the nucleus of comet 67P/C-G bring along further complexity, including surface roughness and boulders, which are not resolved in the typical grid sizes used for torque computations.
We study the dependence of the torque on the mesh size using remeshed versions of the shape model of \cite{Preusker2017} for comet 67P/C-G.
Also for other comets shape models have been derived.
The combination of light curve inversion for convex bodies of \cite{Kaasalainen2001} and radar imaging of \cite{Ostro2002} yield coarse resolution for surface geometry such as in \cite{Donaldson2023} for 162P/Siding Spring and in \cite{Magri2011} for Asteroid (8567) 1996 HW1.
From spacecraft imagery the shapes of comets 1P, 9P, 19P, 67P, 81P, and 103P are constrained \citep[see][]{Snodgrass2022}.
\cite{Knight2024} classify four of these comets as having a bilobed shape.

Observational data for the rotation state of comets are mainly extracted from light curves and radar detections.
The retrieved rotation periods are in the range between a few hours and several days.
The rotation period can change over time as is reported by \cite{Belton2011} and \cite{Chesley2013} for 9P/Tempel 1.
Even the tendency might change from increase to decrease as is reported by \cite{Farnham2021} for comet 46P/Wirtanen during one perihelion passage.
Earth-bound observations constrain the orientation of the rotation axis by studying coma morphology.
Measurements are typically associated with uncertainties on the order of $15^\circ$, see \cite{Knight2021}.
For comet 67P/C-G, orientations of the rotation axis are extracted by \cite{Jorda2016} based on the method of \cite{Gaskell2008}.
Models for the rotation state \citep[see][]{Kramer2019a,Attree2019,Attree2023a,Attree2024} might match complementing quantities, such as the rotation period or the production curve for global gas production.

Gas emanating from the surface carries along momentum determined by the molecular mass and velocity of the volatiles.
Thermophysical models with a monotonous response of the sublimation to the solar irradiation together with the normal direction at each surface point establish the vector fields for forces and torques on the surface.
The torque vector field can be reinterpreted as a weighted superposition of the vectorial torque efficiency.
This quantity depends on the short axis vector and local properties of the shape geometry only.
One scalar component of this torque efficiency that affects the rotation period is introduced by \cite{Keller2015}.
\cite{Jewitt1997} and \cite{Samarasinha2013} consider surface integrated torque terms to reflect cancelation effects.
By integrating spatially the vector field for torque, the resulting torque vector influences the rigid body dynamics.
This depends on the body geometry, on the heliocentric distance, and on the orientation state (the illuminated cross section) of the body relative to the sun.
Already for a monotonous TPM, surface features and self-shadowing can shift the maximum sublimation activity away from the current subsolar point.
Qualitatively, a resulting force caused by irradiation pushes the nucleus into the antisolar direction.
For the torque and changes of the rotation state, such a simple relation does not persist since the shape of the nucleus and the orientation of the rotation axis have to be considered in detail.
For a lowest-energy rotation,
\cite{Kramer2019a} parameterize diurnal torques by the subsolar longitude and derive a torque decomposition as a product of an orbital term and a shape-dependent matrix.
Subsolar latitude, together with solar distance, is associated with seasonal variations.

Although uniform activity on the complex shape of comet 67P/C-G explains observed changes for the rotation period, there is a mismatch between this activity and observations for the axis orientation.
For this problem within a model for comet 67P/C-G, a solution is to prescribe a domain decomposition consisting of surface regions and to fit the corresponding activity weights (effective active fraction).
These weights are only weakly constrained and it is difficult to obtain appropriate decompositions of the surface.
For any cometary body geometry, the vectorial torque efficiency (with its local characterization of the geometry) offers the option to specify a domain decomposition consisting of eight regions.
For lowest-energy rotations and independent of the shape, these regions have the significant property that all points of one region share the same direction for torque formation.

The manuscript describes our approach to decomposing a cometary surface into eight regions that determine the direction of the torque and the resulting evolution of the
rotation state close to lowest energy.
Sect.~\ref{sec:dynamics} introduces the dynamical equations and discusses the driving forces for rotational evolution.
We consider instantaneous TPMs with a weighted activity on the cometary surface.
Sect.~\ref{sec:torque} describes the formation of a torque field at each point on the cometary surface.
The relation between the specified regions and the torque direction is examined for representative examples, namely, for an oblate ellipsoid, a near-prolate ellipsoid, a bilobed shape, and a shape model of comet 67P.
These torques guide the evolution of
the rotation state close to lowest energy in Sect.~\ref{sec:velocity}.
Rotational data for comet 67P/C-G serve as constraints for effective active fractions with respect to our domain decomposition.

\section{Dynamics of rotation}
\label{sec:dynamics}

The rotation state of a rigid body can be described by the temporal evolution of a coordinate mapping $x(\xi,t)$ from the body frame to the inertial frame at time $t$ with a position vector $\xi\in\real^3$ in the body frame.
$x$ is a linear mapping $x(\xi,t) = R\xi$ represented by an orthogonal rotation matrix $R(t)\in\real^{3\times 3}$.
Within the body frame, a tensor of inertia $J_\mathrm{BF}\in\real^{3\times 3}$ is prescribed.
The corresponding tensor is $J(t) = R J_{\mathrm{BF}}R^{-1}$ in the inertial frame.
Based on the vectors for the angular velocity, $\omega(t)\in\real^3$, for the rotational angular momentum, $L(t)\in\real^3$, and the prescribed torque, $T(t)\in\real^3$, the equations for rotation in the inertial frame are
\begin{gather}\label{eq:rotation}
\dot R() = \omega \times R(),\quad
\dot L = T,\quad
J\omega = L.
\end{gather}
Together with the initial conditions, $R(t_0) = R_0$ and $L(t_0)=L_0$, at time $t_0$, this equation poses an initial value problem that determines the temporal evolution of the rotation state, $R(t)$ and $L(t)$.
From the solution of the differential equation, additional variables such as $\omega$, $x(\xi,t)$ are determined.
We take as directions of the inertial frame the J2000 system, with the origin at the center of mass of the cometary body.
Far away from further gravitationally interacting bodies, this translation does not lead to further torques or modifications of Eq.~\eqref{eq:rotation}.
Within this manuscript all results are based on numerical simulations for the full rotational dynamics.

The equation in the inertial frame has an equivalent representation in the body frame with the Euler equations for rotation,
\begin{gather}\label{eq:euler}
J_{\mathrm{BF},ij} \dot \omega_{\mathrm{BF},j}
= T_{\mathrm{BF},i}
- \epsilon_{ijk}\, \omega_{\mathrm{BF},j}
\,J_{\mathrm{BF},kl}\,
\omega_{\mathrm{BF},l},
\end{gather}
for variables $\omega_\mathrm{BF} = R^{-1} \omega$ and $T_\mathrm{BF} = R^{-1} T$.

\subsection{Rigid body geometry}
\label{sec:systems}

Within the body frame with the standard basis vectors, $e_1, e_2, e_3\in\real^3$, the spatial domain $\Omega\subset\real^3$ describes the geometrical shape (body shape) of the cometary nucleus.
In the inertial frame at time $t$ the location of the body is the spatial domain $\Omega_t= x(\Omega,t)$.
We consider three idealized body shapes, $\Omega$, complemented by one additional shape, an approximation of the cometary nucleus of comet 67P/C-G.
The first two shapes are ellipsoids and are oriented such that their symmetry axes coincide with $e_1, e_2, e_3$.
The dimensions of the ellipsoids are characterized by the lengths of the semimajor axis $a_1\geq a_2\geq a_3$ with respect to $e_1, e_2, e_3$.
The first shape is an oblate ellipsoid with $a_1=a_2=2 a_3$ and is therefore axially symmetric around $e_3$.
An oblate ellipsoid underlies the forced precession models discussed in \cite{Whipple1979} and \cite{Sekanina1984}, together with further assumptions on the activity of regional jets on the surface.
The second shape is a near-prolate ellipsoid characterized by $9a_1 = 18 a_2= 20 a_3$ without axial symmetry.
As a consequence of three different axis lengths this shape is not axially symmetric.
That is why we prefer the denotation near-prolate instead of prolate.
For the numerical study here, we describe the boundaries of $\Omega$ for these two shapes using a surface, $\partial\Omega$, consisting of a triangular mesh with $N_\mathrm{E}=4000$ elements.
The third geometry is given by a bilobed shape and consists of two slightly oblate ellipsoids placed side by side without overlap.
Each ellipsoid satisfies the relation $9a_1 = 9a_2= 10 a_3$ such that the bilobed shape holds the same spatial dimensions as the near-prolate ellipsoid.
This bilobed shape configuration could represent a contact binary and is similar to the shape of comet 67P/C-G.
The base vector $e_1$ is located on the line through the centers of the spheres and $e_2, e_3$ span the perpendicular plane.
The surface of the domain, $\Omega$, for the bilobed shape consists of $N_\mathrm{E}=4000$ triangular elements.
The last case is a shape model of comet 67P/C-G.
The $N_\mathrm{E}=3996$ surface elements of the last shape are the same as those of \cite{Lauter2022}, which are derived from the high-resolution (meter-scale) shape model of \cite{Preusker2017}.
All shapes share the same normalized volume $|\Omega| = 18.56\,\mathrm{km}^3$.
Based on these choices, all shapes share the same spatial resolution of approximately $\Delta x\approx 150\,\mathrm{m}$.

The components of the tensor of inertia are given by
\begin{gather}\label{eq:toinertia}
J_{\mathrm{BF},ij} = \intl_{\Omega} \rho
(\delta_{ij} \xi^2 - \xi_i\xi_j ) \,d\xi.
\end{gather}
For the two ellipsoids and for the bilobed shape within domain $\Omega$, we assume the constant density $\rho=539\,\mathrm{kg}/\mathrm{m^3}$. 
For shape 67P, the tensor of inertia $J_{\mathrm{BF}}$ is set to Eq.~\eqref{eq:tensor}.
For this case, the assumption of uniform density is suspended, in good agreement with the gravitational constraints of \cite{Laurent-varin2024}.
For all shapes considered, $e_3$ is the principal axis of $J_\mathrm{BF}$ with the largest eigenvalue, $J_3$.
The vector $e_3$ in the body frame corresponds to the principal axis, $Re_3$, in the inertial frame.
In the case of ellipsoids the moments of inertia are ordered in inverse proportion to the lengths of the semimajor axes.
The principal axis $e_3$ corresponds to the largest moment of inertia and to the shortest axis.
Let us denote $b_3 = Re_3$ as the short axis in the inertial frame.

For a lowest-energy rotation state (represented by $R$ and $L$), the directions of $L$, of $\omega$, and of the short axis coincide.
During the 2015 apparition \cite{Godard2017}, \cite{Preusker2017}, \cite{Jorda2016}, and \cite{Gutierrez2016} consider comet 67P/C-G to be in
lowest-energy rotation.
A free nutation motion for the angular velocity vector has not been observed within the accuracy range of a few tens of millidegrees.
A lowest-energy rotation implies $\omega = |\omega| b_3$ for the angular velocity and $L = J_3 |\omega| b_3$ for the rotational angular momentum.
In the following, the initial states, $R_0$ and $L_0$, are chosen to be at the lowest energy.
With the comet becoming active, the NGTs introduce some perturbations leading to an excited state.
A torque, $T = T_3 b_3$, directed along $b_3$, is able to change the vector length $|\omega|$ (and thus the rotation period), while the directions of $b_3$, $\omega$, and $L$ remain unchanged.
A torque, $T$, that is orthogonal to $b_3$, namely with $T\cdot b_3 = 0$, breaks the initial alignment of $b_3$ and $\omega$ in general.
For linearized Euler equations in Eq.~\eqref{eq:euler}, an analysis for the angle $\gamma$ between $b_3$ and $\omega$ yields the estimation $\tan \gamma \approx |T|/J_3/\omega^2$.
Assuming characteristic values for ice sublimation on comet 67P/C-G with $|T|= 10^9\,\mathrm{Nm}$, $J_3=10^{19}\,\mathrm{kg}\,\mathrm{m^2}$, and $|\omega|=10^{-4}\,\mathrm{s}^{-1}$ yields an upper bound $|\gamma| \approx 0.5^\circ$.

\begin{figure}[t]
\centering
\includegraphics[width=0.3\textwidth]{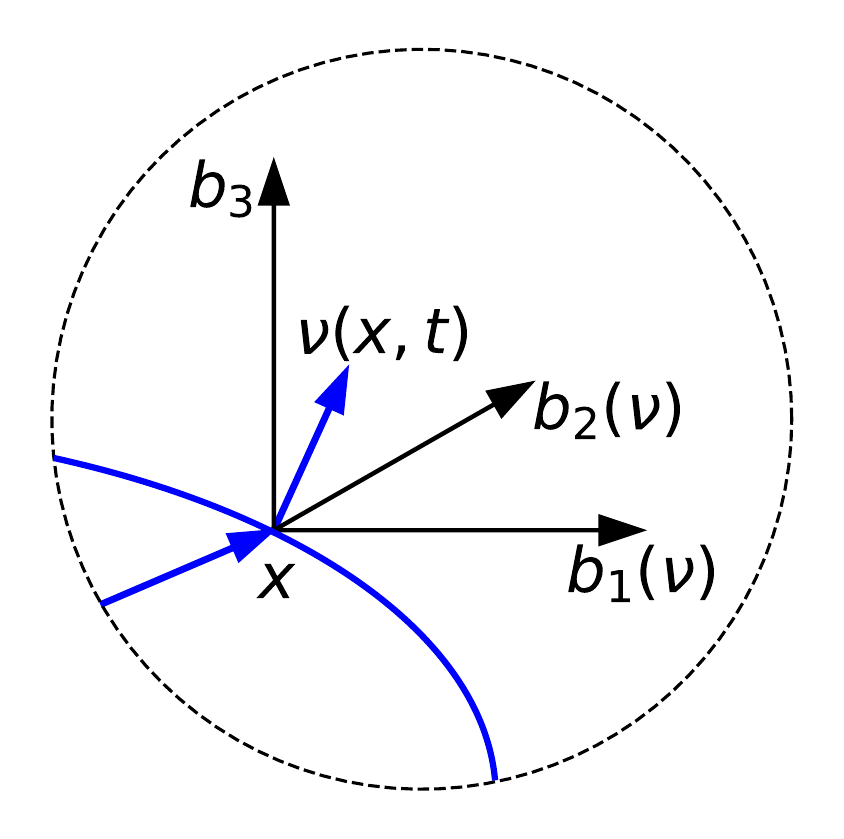}
\caption{Local basis, $B(\nu) = ( b_1(\nu),b_2(\nu),b_3 )$, for the choice $y=\nu(x,t)$ in Eq.~\eqref{eq:sekanina} at time $t$ and at position $x\in\partial\Omega_t$ with the outward normal $\nu$.}
\label{fig:local}
\end{figure}
\begin{figure}[t]
\centering
\includegraphics[width=0.49\textwidth]{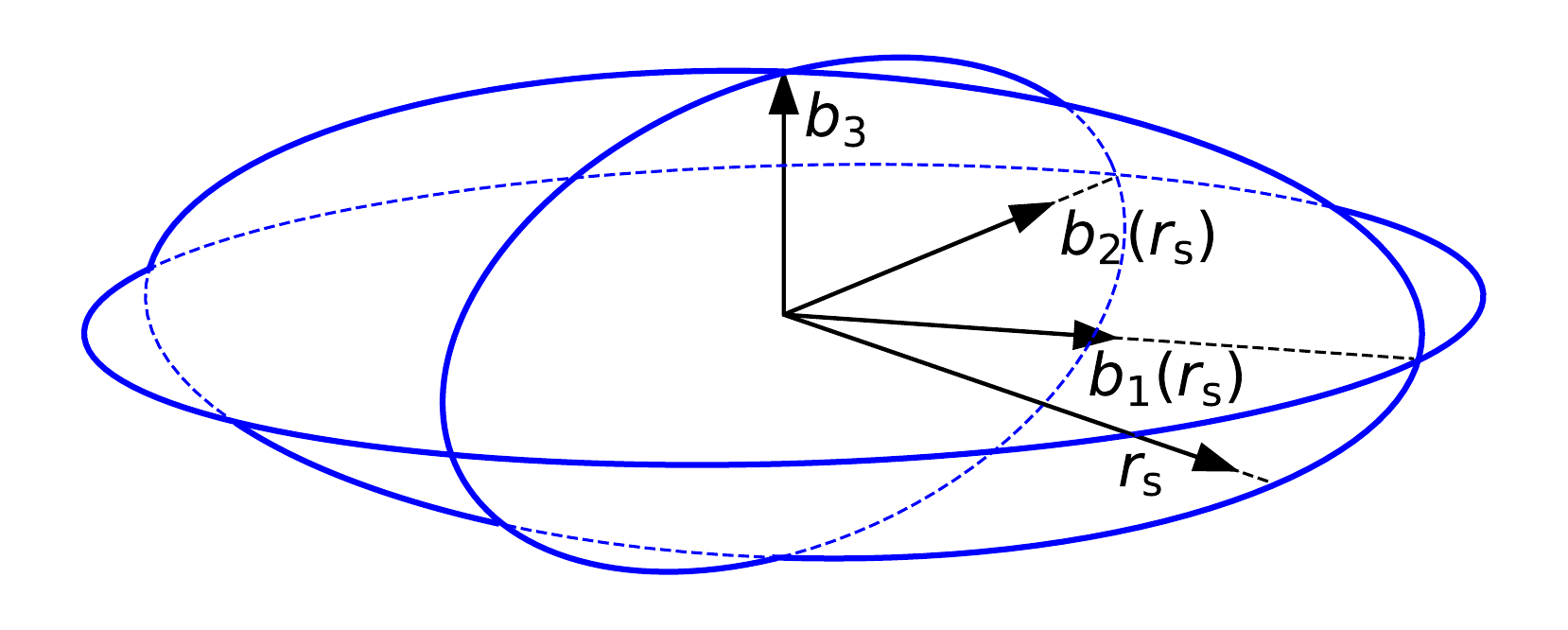}
\caption{Solar basis, $B(r_\mathrm{s})= (b_1(r_\mathrm{s}),b_2(r_\mathrm{s}),b_3)$,  for the choice $y=r_\mathrm{s}(t)$ in Eq.~\eqref{eq:sekanina} at time $t$.}
\label{fig:solar}
\end{figure}
To analyze the orientation of vector quantities (e.g., torque and angular velocity) relative to the sun and to the short axis, we defined two different basis systems with the (orthogonal) vectors
\begin{equation}\label{eq:sekanina}
\begin{gathered}
b_1(y) =
\frac{(b_3\times y)\times b_3 }{|(b_3\times y)\times b_3|},\quad
b_2(y) =
\frac{b_3\times y}{|b_3\times y|},
\quad
b_3 = R e_3.
\end{gathered}
\end{equation}
We chose the vector $y\in\real^3$ in two different variants that yield the local basis and the solar basis, respectively.
As shorter denotation we also use $B(y) = (b_1(y), b_2(y), b_3 )$.

For the local basis at time $t$ and a position vector $x\in\partial\Omega_t$ on the surface, we defined $y=\nu(x,t)$.
Fig.~\ref{fig:local} shows the outward normal vector $\nu$ on $\partial\Omega_t$.
At a fixed time $t$ the local basis, $B(\nu)$, depends on the surface position $x$ and on the principal axis $b_3$.
With advancing time $t$ at a fixed position $\xi\in\partial\Omega$, this basis in the body frame (namely $R^{-1}b_1(\nu), R^{-1}b_2(\nu), R^{-1}b_3$) is constant.
This means that the local basis follows the rotational movement in the inertial frame together with $x=R\xi$.
The three vectors $b_1$, $\nu$, and $b_3$ are located in a shared plane.
For the solar basis, $B(r_\mathrm{s})$, in Eq.~\eqref{eq:sekanina} we define $y=r_\mathrm{s}(t)$ with the vector $r_\mathrm{s}$ pointing toward the sun at time $t$.
Fig.~\ref{fig:solar} shows this geometric setup.
At a fixed time $t$ the basis is constant along the surface, $\partial\Omega_t$.
With advancing time $t$, it does not follow the rotational movement in the inertial frame and it rotates in the body frame.
The vector $b_1$ is related to the solar direction; that is, such that $b_1$, $r_\mathrm{s}$, and $b_3$ are located in a shared plane.
Except for the signs for a lowest-energy state, this solar basis was already applied by \cite{Kramer2019} and \cite{Sekanina1981}.
\cite{Kramer2019a} apply this basis to represent the diurnally averaged torque affecting the rotation state.
The local and solar basis share the property that $b_3$ is aligned to the short axis so that the equatorial plane of the rotation is spanned by the vectors $b_1$ and $b_2$.
The equatorial planes of both basis systems are the same.
$b_1(\nu), b_2(\nu)$ and $b_1(r_\mathrm{s}), b_2(r_\mathrm{s})$ are just two choices for a basis of this plane.

\subsection{Domain decomposition of the surface}

For each $\xi\in\partial\Omega$ on the surface, $x = R\xi$, $\nu(x,t)$, and $b_i(\nu)$ (the local basis vector) follow the same rotational movement in the inertial frame.
Based on these quantities, we define regions on $\partial\Omega_t$ that move correspondingly in the inertial frame.
Their counterparts in the body frame are time-independent, so these definitions are based on geometry properties of the shape in the body frame, $\partial\Omega$.

\begin{figure}[t]
\begin{minipage}{0.32\textwidth}
\centering
\includegraphics[width=0.98\textwidth]{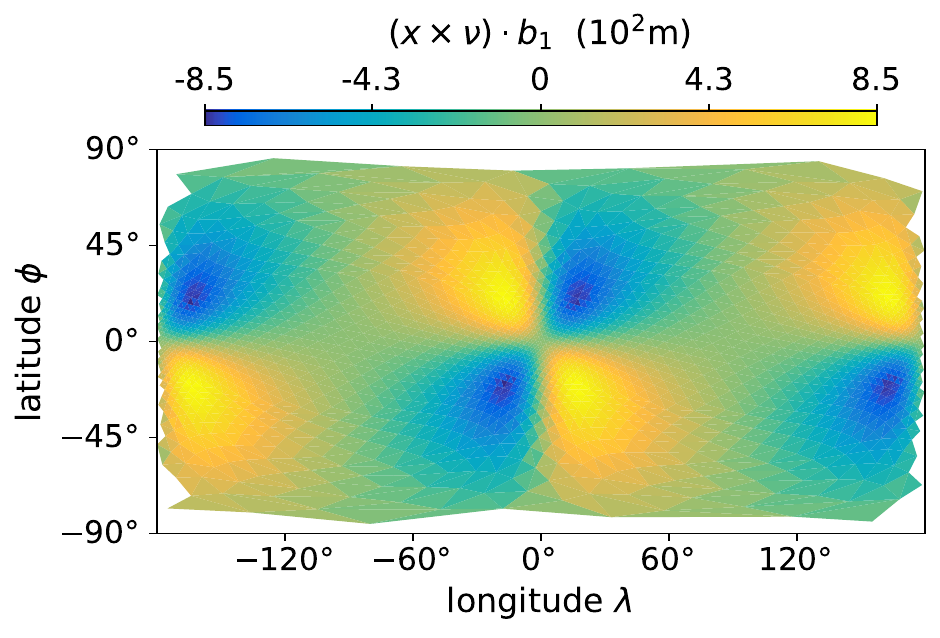}
\end{minipage}
\begin{minipage}{0.32\textwidth}
\centering
\includegraphics[width=0.98\textwidth]{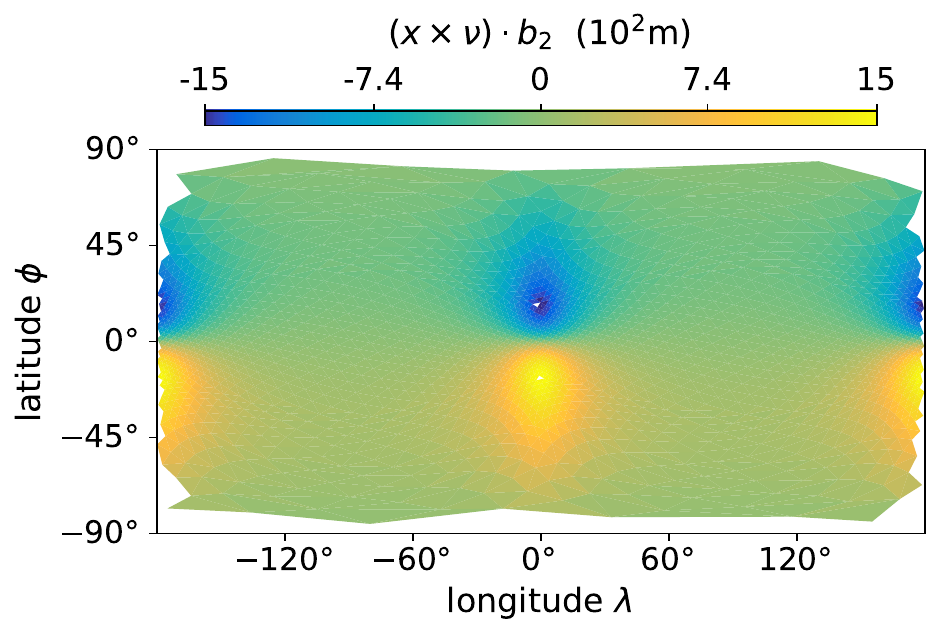}
\end{minipage}
\begin{minipage}{0.32\textwidth}
\centering
\includegraphics[width=0.98\textwidth]{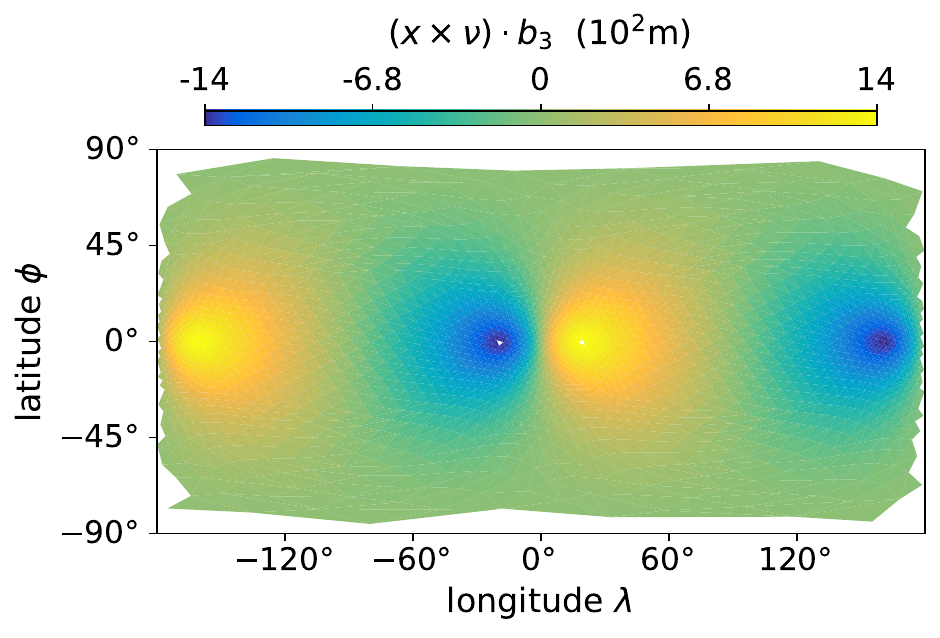}
\end{minipage}
\caption{For the near-prolate ellipsoid, the values $(x\times\nu)\cdot b_{i}(\nu)$ (vectorial torque efficiency) are shown based on longitude, $\lambda$, and latitude, $\phi$, in the body frame.
$b_i(\nu)$ is the local basis vector in Fig.~\ref{fig:local}.
}
\label{fig:projec}
\end{figure}
\begin{figure}[t]
\begin{minipage}{0.48\textwidth}
\centering
\includegraphics[width=0.98\textwidth]{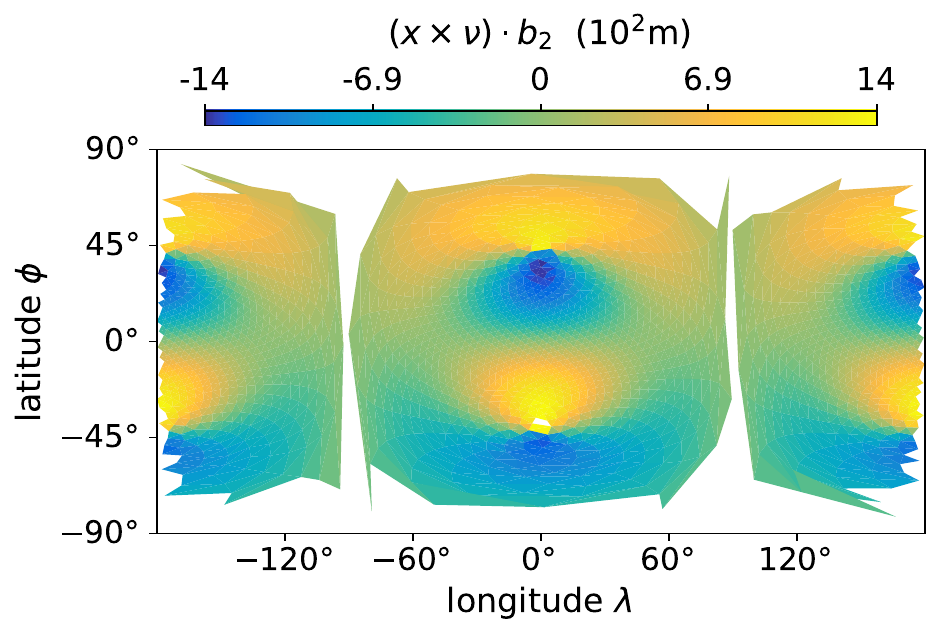}
\end{minipage}
\begin{minipage}{0.48\textwidth}
\centering
\includegraphics[width=0.98\textwidth]{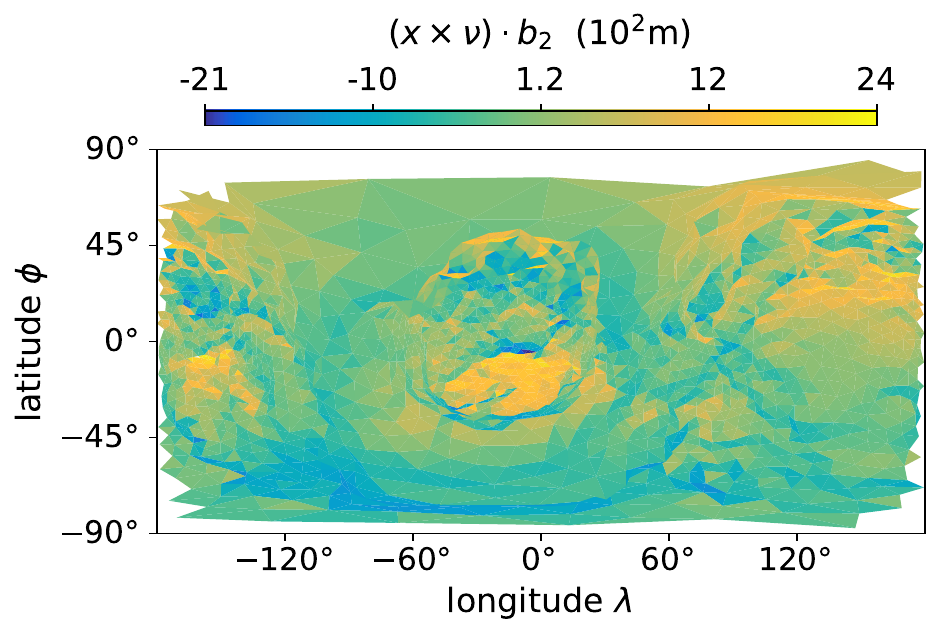}
\end{minipage}
\caption{%
For the bilobed shape (top) and for shape 67P (bottom), the values $(x\times\nu)\cdot b_{2}(\nu)$ (second component of the vectorial torque efficiency) are shown
based on longitude, $\lambda$, and latitude, $\phi$, in the body frame.
$b_2(\nu)$ is the local basis vector in Fig.~\ref{fig:local}.
}
\label{fig:projbilobed}
\end{figure}
At time $t$ we define the southern and the northern hemisphere,
\begin{equation*}
\begin{gathered}
A_\mathrm{SH}
= \{ x \in \partial\Omega_t\,|\, \nu\cdot b_{3} < 0 \}
,\quad
A_\mathrm{NH}
= \{ x \in \partial\Omega_t\,|\, 0 < \nu\cdot b_{3} \}.
\end{gathered}
\end{equation*}
Their specification is based on the sign of the $b_3$ component for the surface normal, $\nu$, at point $x\in\partial\Omega_t$.
For a lowest-energy rotation state, this projection controls the seasonal intensity of the solar irradiation on the surface, $\partial\Omega_t$.
For negative subsolar latitude SSL = $\arcsin(r_\mathrm{s}\cdot b_3/|r_\mathrm{s}|)$, the point of maximum irradiation is located in $A_\mathrm{SH}$.
For positive SSL, this point is located in $A_\mathrm{NH}$.
This relation expresses the mapping between $A_\mathrm{SH}$ and $A_\mathrm{NH}$ on the one side and both hemispheric seasons on the other side.

\begin{figure}[t]
\begin{minipage}{0.48\textwidth}
\centering
\includegraphics[width=0.85\textwidth]{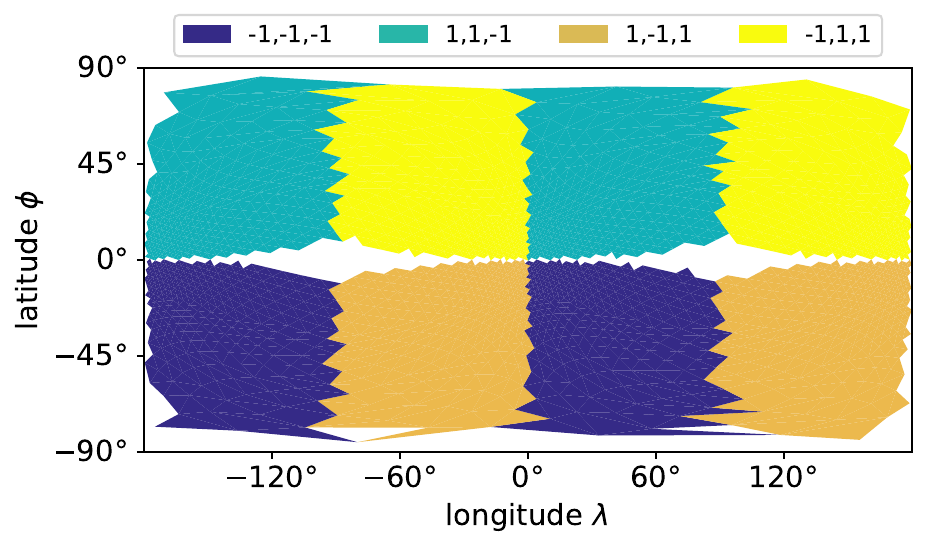}
\end{minipage}
\begin{minipage}{0.48\textwidth}
\centering
\includegraphics[width=0.85\textwidth]{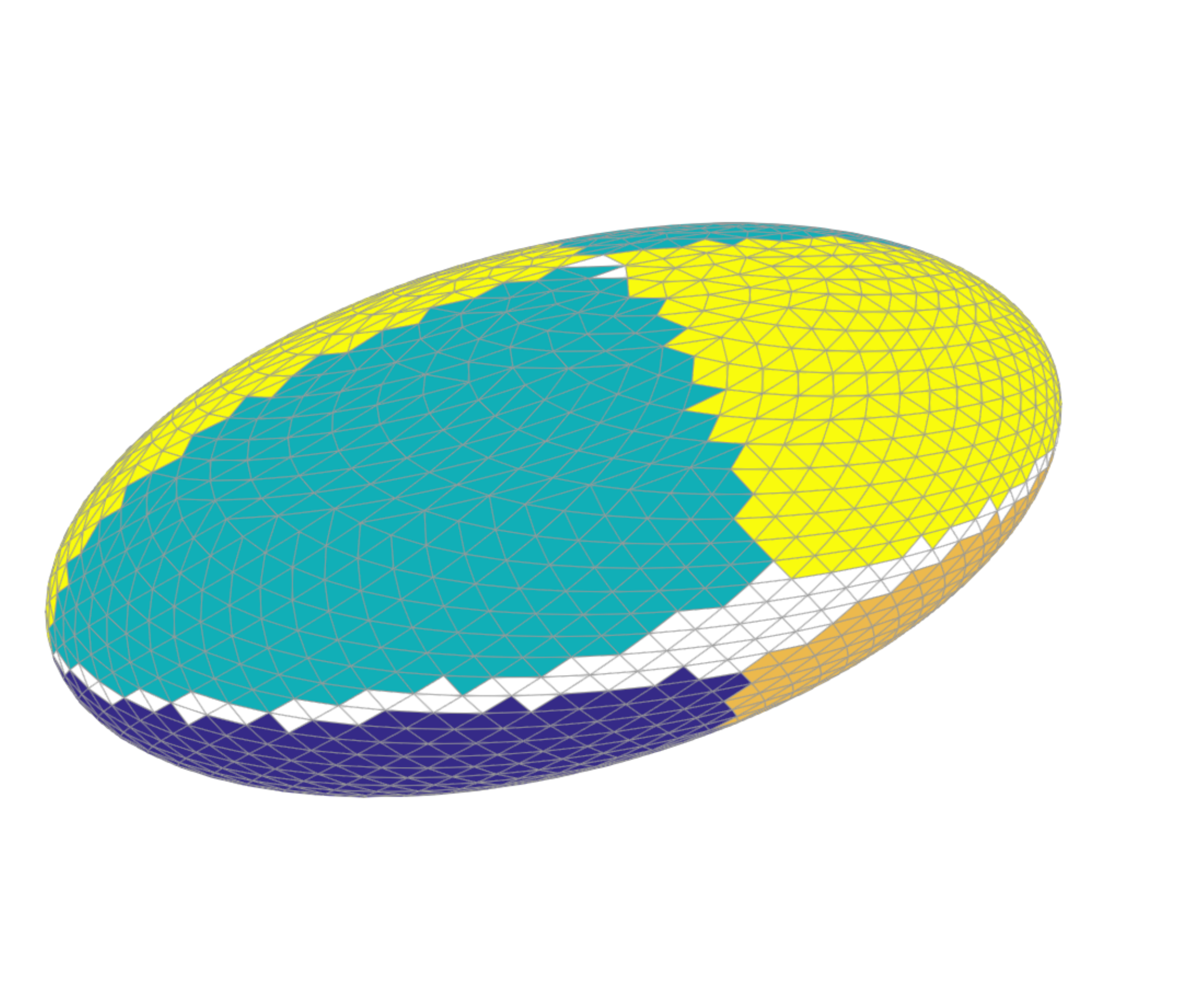}
\end{minipage}
\caption{For the near-prolate ellipsoid decomposition of $\partial\Omega_t$ into the four regions $A_{i,j,k}$ in Eq.~\eqref{eq:regions}.
Region $A_0$ is colored in white.
Top: Longitude, $\lambda$, and latitude, $\phi$, in the body frame.
Bottom: 3D projection using the same color code.}
\label{fig:regions}
\end{figure}
This concept of defining regions based on projections to the local basis in Fig.~\ref{fig:local} can be applied to the vector $x\times\nu$.
We define the vectorial torque efficiency by the three scalar components,
\begin{equation}\label{eq:vte}
(x\times\nu)\cdot b_{i},
\end{equation}
for $i=1,2,3$.
Fig.~\ref{fig:projec} shows the maps of the vectorial torque efficiency for the near-prolate ellipsoid.
Fig.~\ref{fig:projbilobed} shows the $b_{2}$ component for the more complex geometries of the bilobed shape and shape 67P.
In Sect.~\ref{sec:torque} we show that both figures represent components of the vectorial torque efficiency.
Based on these fields for $i,j,k = \pm 1$, we define eight regions,
\begin{equation}\label{eq:regions}
\begin{gathered}
A_{i,j,k} =
\left\{ x \in\partial\Omega_t \,\left|\,
\begin{matrix}
i\, (x\times\nu) \cdot b_{1} < 0, \\
j\, (x\times\nu) \cdot b_{2} < 0, \\
k\, (x\times\nu) \cdot b_{3} <0
\end{matrix}
\right.
\right\},
\end{gathered}
\end{equation}
and a complementing zero region,
\begin{gather*}
A_{0} =
\left\{ x \in\partial\Omega_t \,\left|\,
\begin{matrix}
(x\times\nu) \cdot b_{1} = 0 \,\text{or} \\
(x\times\nu) \cdot b_{2} = 0 \,\text{or} \\
(x\times\nu) \cdot b_{3} = 0
\end{matrix}
\right.
\right\}.
\end{gather*}
The analysis of signs in the three plots of Fig.~\ref{fig:projec} yields these regions in Fig.~\ref{fig:regions} for the near-prolate ellipsoid.
In this case, the regions $A_{1,-1,-1}$, $A_{-1,1,-1}$, $A_{-1,-1,1}$, and $A_{1,1,1}$ are empty, which yields the decomposition $\partial\Omega_t = A_0\cup A_{-1,-1,-1} \cup A_{1,1,-1}\cup A_{1,-1,1}\cup A_{-1,1,1}$ into four regions and $A_0$.
For the oblate ellipsoid, the rotational symmetry around $b_{3}$ yields $(x\times\nu)\cdot b_{1} = (x\times\nu)\cdot b_{3} = 0$ and finally $\partial\Omega_t = A_0$.
These two examples show that symmetries can lead to empty regions $A_{i,j,k}$.

\begin{figure}[t]
\begin{minipage}{0.48\textwidth}
\centering
\includegraphics[width=0.98\textwidth]{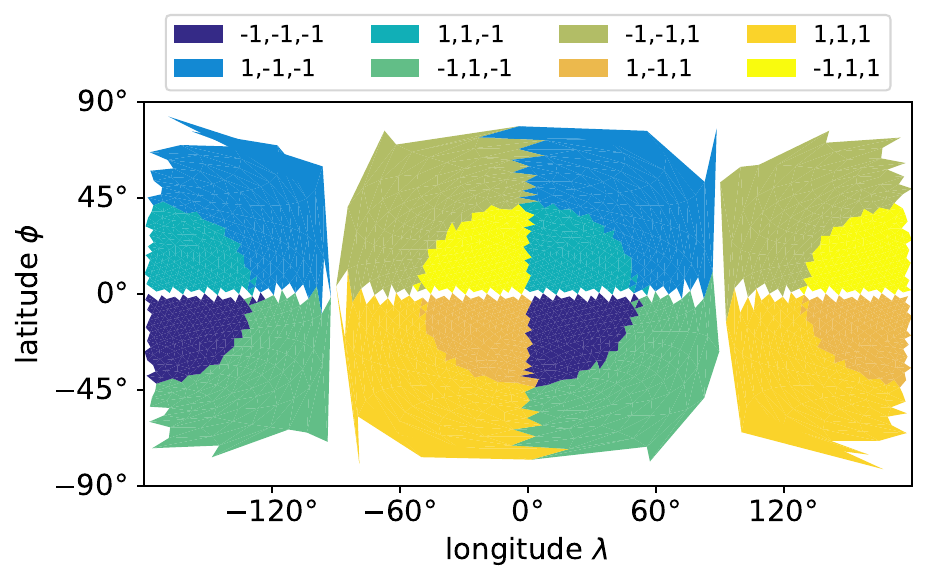}
\end{minipage}
\begin{minipage}{0.48\textwidth}
\centering
\includegraphics[width=0.98\textwidth]{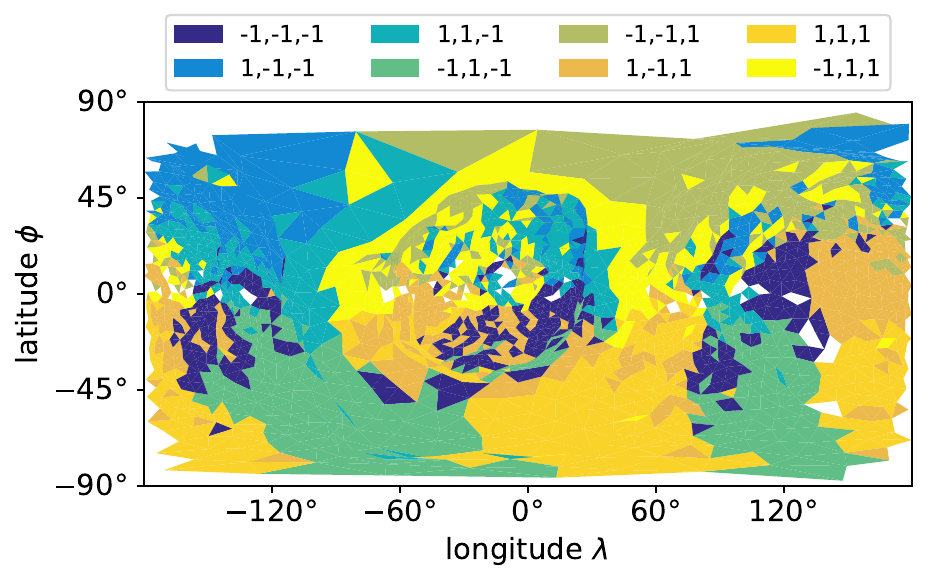}
\end{minipage}
\caption{For the bilobed shape (top) and for the shape 67P (bottom),
 decomposition of $\partial\Omega_t$ into the eight regions $A_{i,j,k}$ in Eq.~\eqref{eq:regions}.
Region $A_0$ is colored in white.
Longitude, $\lambda$, and latitude, $\phi$, in the body frame.}
\label{fig:regbilobed}
\end{figure}
\begin{figure}[t]
\begin{minipage}{0.48\textwidth}
\centering
\includegraphics[width=0.98\textwidth]{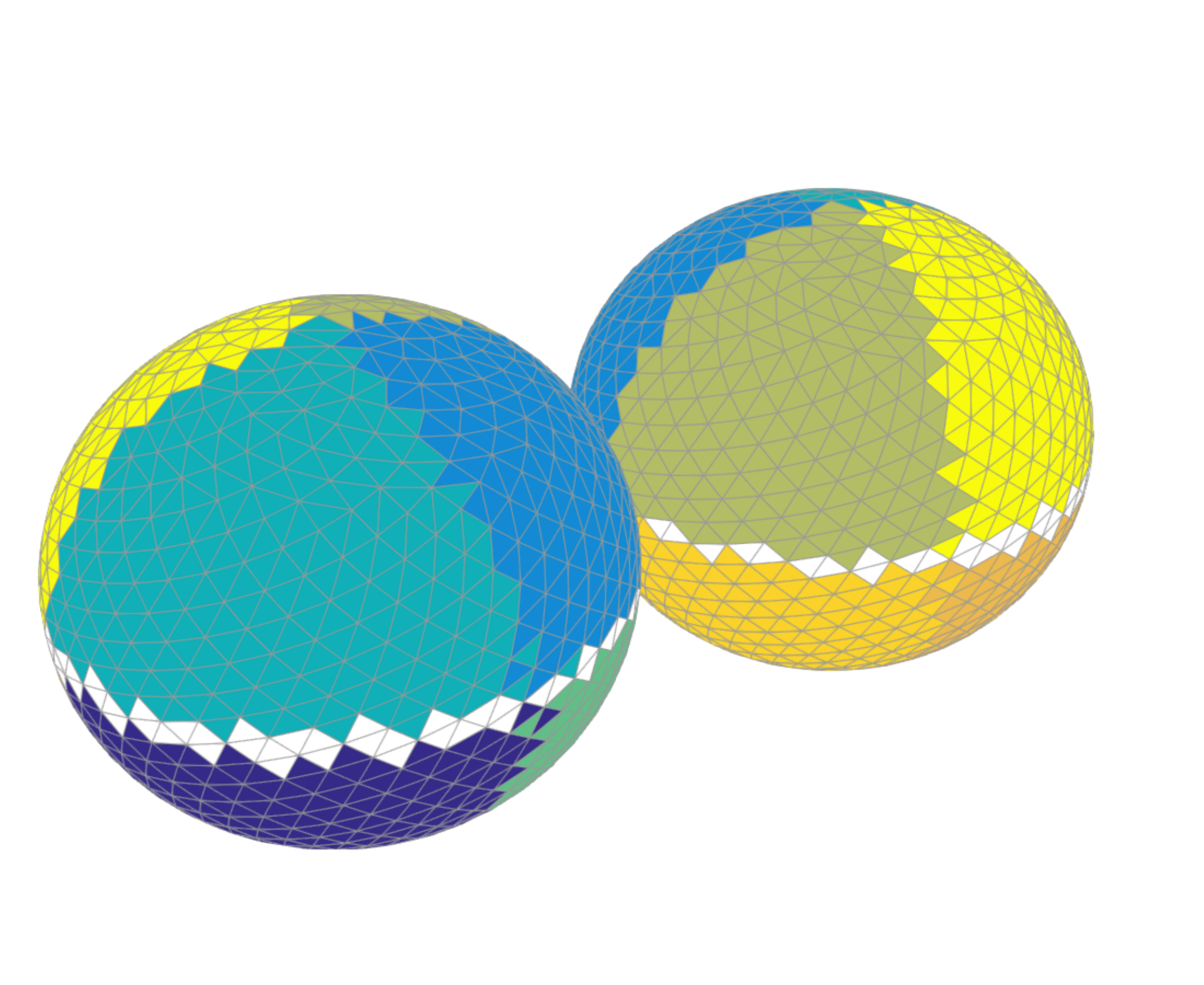}
\end{minipage}
\begin{minipage}{0.48\textwidth}
\centering
\includegraphics[width=0.98\textwidth]{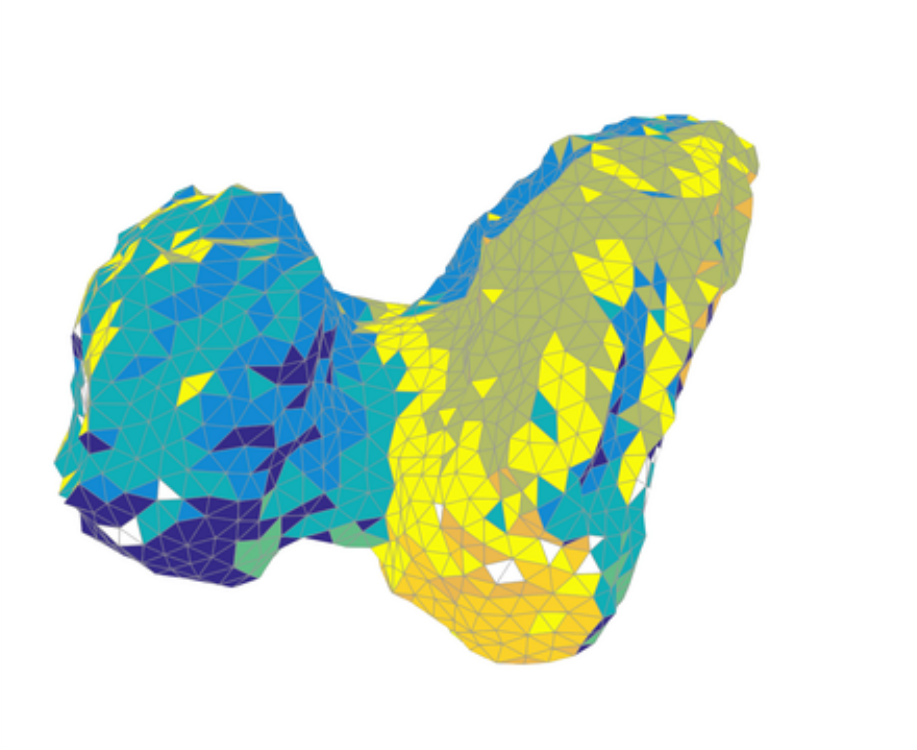}
\end{minipage}
\caption{For the bilobed shape (top) and for the shape 67P (bottom),
decomposition of $\partial\Omega_t$ into the eight regions.
The same data as in Fig.~\ref{fig:regbilobed} but in 3D projection.}
\label{fig:reg3d}
\end{figure}
In general, all eight surface regions, $A_{i,j,k}$, in Eq.~\eqref{eq:regions} are nonempty, as shown in Figs.~\ref{fig:regbilobed}, \ref{fig:reg3d} for the bilobed shape and for the shape 67P.
Applying Eq.~\eqref{eq:oneandthree}, each region is a subset of $A_\mathrm{SH}$ or $A_\mathrm{NH}$ at most, and the following decompositions hold.
\begin{equation}\label{eq:hemirelation}
\begin{gathered}
A_\mathrm{SH}  =
A_{-1,\pm 1,-1} \cup
A_{1,\pm 1,1} \cup A_0
,\\
A_\mathrm{NH} =
A_{-1,\pm 1,1} \cup
A_{1,\pm 1,-1} \cup A_0.
\end{gathered}
\end{equation}
Both properties can be recognized in Figs.~\ref{fig:regions}, \ref{fig:regbilobed}, and \ref{fig:reg3d}.
Looking straight from the south or from the north, only four of the eight regions are visible.
For a lowest-energy rotation state, Eq.~\eqref{eq:hemirelation} expresses that each hemispheric season is related to the predominant irradiation of one of both subsets.

Based on ice sublimation, torque creation will be discussed in the next section.
For that, regionally differing weights for activity will be considered.
This is realized by the function $f_\mathrm{EAF}$ for the effective active fraction on the surface, $\partial\Omega_t$.
The uniform activity is represented by the constant function $f_\mathrm{EAF} = 1$.
Based on the decomposition of $\partial\Omega_t$ into the regions $A_0$ and $A_{i,j,k}$, the effective active fraction{} is defined by
\begin{gather}\label{eq:super}
f_\mathrm{EAF} = \sum_{i,j,k=\pm 1} f_{\mathrm{EAF},i,j,k} \chi_{A_{i,j,k}}.
\end{gather}
$\chi_{A_{i,j,k}}$ is the characteristic function on the region $A_{i,j,k}$.
The weights $f_{\mathrm{EAF},i,j,k}$ are open parameters for $f_\mathrm{EAF}$, which need to be restricted by the observational data.

\subsection{Thermophysical models}
\label{sec:tpm}

The torque, $T$, in Eq.~\eqref{eq:rotation} modifies the rotational angular momentum, $L$, directly and, as a consequence, controls the evolution of the full rotation state.
On the nucleus, $\Omega_t$, solar irradiation causes ice sublimation close to the surface, which leads to gas expansion into the surrounding space.
On a surface point, we denote as a TPM the mappings from solar irradiation, $I$, (and its temporal variation) to mass generation (due to sublimation), $\dot m$, and to gas velocity, $v_\mathrm{gas}$.
Thermophysical models consider complex dynamical process for material, energy and momentum on the surface and within the volume of the nucleus below.
Interactions with subsurface layers can introduce thermal inertia, which causes changes in $I$ to appear for $\dot m$ and $v_\mathrm{gas}$ with a temporal delay.

For the local evaluation of solar irradiation, $I$, we take into account the shadowing of surface areas due to the concave shape of $\Omega_t$.
If there is no additional intersection between the line of sight from the surface position $x\in\partial\Omega_t$ to the sun at position $r_\mathrm{s}$, the solar irradiation is given by
\begin{gather}\label{eq:idef}
I(x,r_\mathrm{s}) = I_\mathrm{s}
\left(\frac{1\mathrm{au}}{|r_\mathrm{s}|}\right)^2
\frac{\nu\cdot r_\mathrm{s}}{|r_\mathrm{s}|}
\end{gather}
with the solar constant $I_\mathrm{s}$.
In the following, we restrict ourselves to instantaneous TPMs without thermal inertia.
This means that both functions $\dot m(I)$ and $v_\mathrm{gas}(I)$ are functions of solar irradiation, $I$, only.
Another assumption for our TPMs is that both functions $\dot m(I)$ and $v_\mathrm{gas}(I)$ increase monotonously with respect to $I$.

For quantification, we consider two specific choices for instantaneous TPMs, model A and model $\alpha$.
Model A is based on the instantaneous energy balance of a irradiated surface covered by pure ice.
Our model A is the TPM for water of \cite{Lauter2022}.
For less intense irradiation, $I$, of up to 50~$\mathrm{W/m}^2$ the mass production is close to zero.
For higher intensities in the range of up to 1000~$\mathrm{W/m}^2$ this functional relation is close to a linear relation.
The second TPM model $\alpha$ with the functions $\dot m_\alpha(I)$ and $v_\alpha(I)$ has been introduced by \cite{Kramer2019a}.
This approach goes back to the observation during the 2015 perihelion passage of comet 67P/C-G, that the gas production increases more rapidly than predicted from model A.
For an exponent $1\leq \alpha$ we defined the TPM by
\begin{gather*}
\dot m_\alpha(I)
= \dot m_0
\left(\frac{I}{I_0}\right)^\alpha
,\quad
v_\alpha = 450\, \mathrm{m/s}.
\end{gather*}
The constant $\dot m_0 = 3.47\times 10^{-4}\,\mathrm{kg/s/m}^2$ was chosen so that $\dot m_\alpha$ and model~A yield the same sublimation rate at irradiation $I_0 = 1000\, \mathrm{W/m}^2$.
Although any choice for $\alpha$ is possible, we restrict our consideration to the value $\alpha = 2$.

Assuming the gravitational acceleration acting on the body is constant along the body domain, the torque caused by gravitation is zero.
As a consequence, the torque vector, $T(t)$, is restricted to nongravitational contributions at time $t$.
Based on an instantaneous TPM, gas expansion starts from each point $x$ on the surface, $\partial\Omega_t$.
The torque, $T$, and the vector field, $t_\mathrm{NG}$, for the torque density on $\partial\Omega_t$ are defined by
\begin{equation}\label{eq:ngt}
\begin{gathered}
T(t) = \intl_{\partial\Omega_t} t_\mathrm{NG} \,d\sigma
,\quad
t_\mathrm{NG}(x,t)
= -  f_\mathrm{EAF}\, v_\mathrm{gas}\, \dot m\, x \times \nu.
\end{gathered}
\end{equation}
Because the velocity $\omega\times x$ (related to the angular velocity, $\omega$) is small compared to $v_\mathrm{gas} \nu$, we have omitted this fraction of the velocity.
The momentum transfer coefficient applied by \cite{Rickman1989}, \cite{Gutierrez2005}, and \cite{Attree2019} is also omitted.
Because it is a constant we include it in $f_\mathrm{EAF}$, which is still an open parameter.
Eq.~\eqref{eq:ngt} expresses a relation between the intensity of solar irradiation and the strength of the torque.
In particular, each of both hemispheric seasons is now related to torque activity in one of the two separate subsets in Eq.~\eqref{eq:hemirelation}.

Instantaneous TPMs create torque without delay in time.
That is why the solar direction of the vector $r_\mathrm{s}$ and with it the vector $b_1(r_\mathrm{s})$ are associated with the preferred direction of forces on the surface.
For rotational dynamics, these forces lead to a preferred torque direction in Eq.~\eqref{eq:ngt}
associated with $\pm b_2(r_\mathrm{s})$ orthogonal to the solar direction.
Any vector $v\in\real^3$ can be represented by the three components $v\cdot b_i(r_\mathrm{s})$.
With these components, longitude, $\lambda_\mathrm{s}$, and latitude, $\phi_\mathrm{s}$, of $v$ with respect to the solar basis are defined by $\lambda_\mathrm{s} = \arctan(v\cdot b_2(r_\mathrm{s})/v\cdot b_1(r_\mathrm{s}))$ and $\phi_\mathrm{s} = \arcsin(v\cdot b_3/|v|)$.
The other way around, knowing the solar basis, both angles $\lambda_\mathrm{s}$ and $\phi_\mathrm{s}$ determine the direction of $v$.
The case longitude $\lambda_\mathrm{s}\in\{ 0^\circ, 180^\circ \}$ corresponds to $v\cdot b_2 = 0$.
In this case, the solar direction $r_\mathrm{s}$ and $v$ are closely related in the way that both projections to the equatorial plane coincide.
The case $\lambda_\mathrm{s}\in \{ -90^\circ, 90^\circ \}$ corresponds to $v\cdot b_1 = 0$ such that the equatorial projections of $r_\mathrm{s}$ and $v$ are perpendicular.
For latitude $\phi_\mathrm{s} = 0^\circ$, $v\cdot b_3 = 0$ holds and $v$ is in the equatorial plane.
Finally, $\phi_\mathrm{s} \in \{ -90^\circ, 90^\circ \}$ gives $v\cdot b_1 = v\cdot b_2 = 0$ and the representation $v = \pm |v| b_3$.
For the special case of $v=r_\mathrm{s}$, $\phi_\mathrm{s}$ is the same as SSL.

\subsection{Data for the rotation state of 67P/C-G}
\label{sec:data}

For a numerical solution of the initial value problem of Eq.~\eqref{eq:rotation} a set of parameters needs to be specified.
These include the initial conditions, the tensor of inertia, and the mechanism for torque formation in Eq.~\eqref{eq:ngt}.
The solution of the inverse problem consists of the determination of the open parameters based on known observational data.
Angular velocity, $\omega$, was observed during the ESA mission of comet 67P/C-G during the apparition in 2015.
This rotational information refers to the angular frequency, $f=|\omega|$, of the rigid body, as well as to the directional angles $\ra(\omega)$, $\dec(\omega)$ of the rotation axis, $\omega/|\omega|$.
Our data set is based on the measurements for $\omega$ discussed in \cite{Kramer2019a}.
In the equatorial frame J2000 (EME2000) the angles are extracted based on the method of \cite{Gaskell2008} applied by \cite{Jorda2016} to comet 67P/C-G. 
Data sets $\data(\omega) = (\ra(\omega),\dec(\omega),|\omega| )$ are averaged along a moving time interval of 21~days and are considered between August 2014 and July 2016.
Time sampling defines $N_\mathrm{time} = 58$ time points $t_i$ with $i=1,...,N_\mathrm{time}$.
The sampling rate varies between a 7~day to a 33~day distance between two times with a higher sampling rate close to perihelion passage in August 2015.
The uncertainty of the data at time $t$ is estimated based on the spread of the data close to that time.
This leads to uncertainties between $0.01^\circ$ and $0.11^\circ$ for $\dec(\omega)$ and between 11~s and 60~s for rotation period $2\pi/f$.
The latter range is much higher compared to uncertainties on the order of 1~s for periods from \cite{Godard2017} and \cite{Mottola2014}.
We chose these higher weights to distribute the uncertainties uniformly in the direction and in the length of the vector, $\omega$.
We combined these preparations into the data set,
\begin{gather}\label{eq:data}
( t_i,data_{i},\sigma_{i} )
\quad\text{for}\quad
i=1,...,N_\mathrm{time},
\end{gather}
of $3N_\mathrm{time}$ data points $data_{i} = \data(\omega_{i})\in\real^3$ and uncertainties $\sigma_{i}\in\real^3$ at time $t_i$.

\section{Torque formation for lowest-energy rotation}
\label{sec:torque}

The torque, $T$, in Eq.~\eqref{eq:ngt} depends on the geometric properties of the shape $\Omega_t$, on the TPM, and on the weights of the effective active fraction $f_\mathrm{EAF}$ in Eq.~\eqref{eq:super}. 
The sensitivity of $T$ to $\Omega_t$ and $f_\mathrm{EAF}$ is discussed in the present section.
$\Omega_t$ specifies the orientation of the surface regions and, along with it, the formation of the torque density on the surface.
The uniform activity with $f_\mathrm{EAF} = 1$ represents the dynamics of the complete shape.
The combination of spatially weighted effective active fractions in Eq.~\eqref{eq:super} and complex shape geometries gives options to modify the rotational dynamics.
Within this section we restrict the discussion to lowest-energy rotation states.
As a consequence, two significant vectors share the same direction, the rotation axis, $\omega$, and the principal axis, $Re_3$ (the same as $b_3$ in Eq.~\eqref{eq:sekanina}).

The discussion of parameters acting on torque generation is based on the torque components, $T\cdot b_i(r_\mathrm{s})$, with respect to the solar basis in Fig.~\ref{fig:solar}.
Because torque, $T(t)$, in Eq.~\eqref{eq:rotation} determines the derivative $\dot L$, the dynamical effect of $T$ is represented by its integral or its average $\overline{T}$ over time.
Due to the rotation of point $\xi\in\partial\Omega$ with outward normal $\nu_\xi$ in the body frame, the surface point $x(t)=R(t)\xi\in\partial\Omega_t$, the outward normal $\nu(t) = R(t)\nu_\xi$, the irradiation, $I\sim r_\mathrm{s}\cdot\nu(t)$, the gas velocity, $v_\mathrm{gas}$, and the mass production, $\dot m$, change in time.
Consequently, in general, the torque, $T$, is not constant in time.
In particular this is true for the near-prolate ellipsoid and the bilobed shape holding fewer rotational symmetry than the oblate ellipsoid.
At time $t$ for $T$ and $t_\mathrm{NG}$, the temporal averages over one rotation period $\Delta t$ can be expressed by
\begin{equation}\label{eq:taverage}
\begin{gathered}
\overline{T}(t)
= \intl_{\partial\Omega_t} \overline{t_\mathrm{NG}} \,d\sigma
,\quad
\overline{t_\mathrm{NG}}(x,t)
= \frac{1}{\Delta t}\intl_t^{t+\Delta t} t_\mathrm{NG}(x(s),s)\,ds.
\end{gathered}
\end{equation}
The average of $t_\mathrm{NG}$ is evaluated along the trajectory $x(s)\in\partial\Omega_s$ corresponding to $\xi = R^{-1}(t)x$ and $\nu_\xi = R^{-1}(t)\nu$.
Considering constant $\omega$ and $r_\mathrm{s}$, Eq.~\eqref{eq:fouriertrafo} replaces the term $\overline{t_\mathrm{NG}}(x,t)$, which gives
\begin{equation}\label{eq:localt}
\begin{gathered}
T_i = b_{i}(r_\mathrm{s}) \cdot \overline{T} =
 - \intl_{\partial\Omega_t} 
f_\mathrm{EAF} \, c_i\, b_i(\nu)\cdot (x\times\nu) \,d\sigma.
\end{gathered}
\end{equation}
As is described in appendix \ref{sec:fourier}, the numbers $c_i$ are Fourier coefficients for the source term $v_\mathrm{gas}\dot m$ representing the TPM.
A similar idea has been introduced by \cite{Kramer2019a} for the diurnal and seasonal sublimation cycles.
There, three Fourier components are derived for each surface region that produce the torque movement.
On the left-hand side of Eq.~\eqref{eq:localt} are the components of torque, $\overline{T}$, with respect to the solar basis, $B(r_\mathrm{s})$, in Fig.~\ref{fig:solar}.
These represent directions toward and perpendicular to the sun as described in Sect.~\ref{sec:tpm}.
On the right-hand side, $T_i$ is constructed by components with respect to the local basis, $B(\nu)$, in Fig.~\ref{fig:local}.
Adjacent to the TPM terms, the integrand in Eq.~\eqref{eq:localt} contains the
vectorial torque efficiency, $(x\times\nu) \cdot b_i(\nu)$, defined in Eq.~\eqref{eq:vte}.

\subsection{Idealized shape geometries}
\label{sec:simpleshape}

\begin{figure}[t]
\centering
\includegraphics[width=0.45\textwidth]{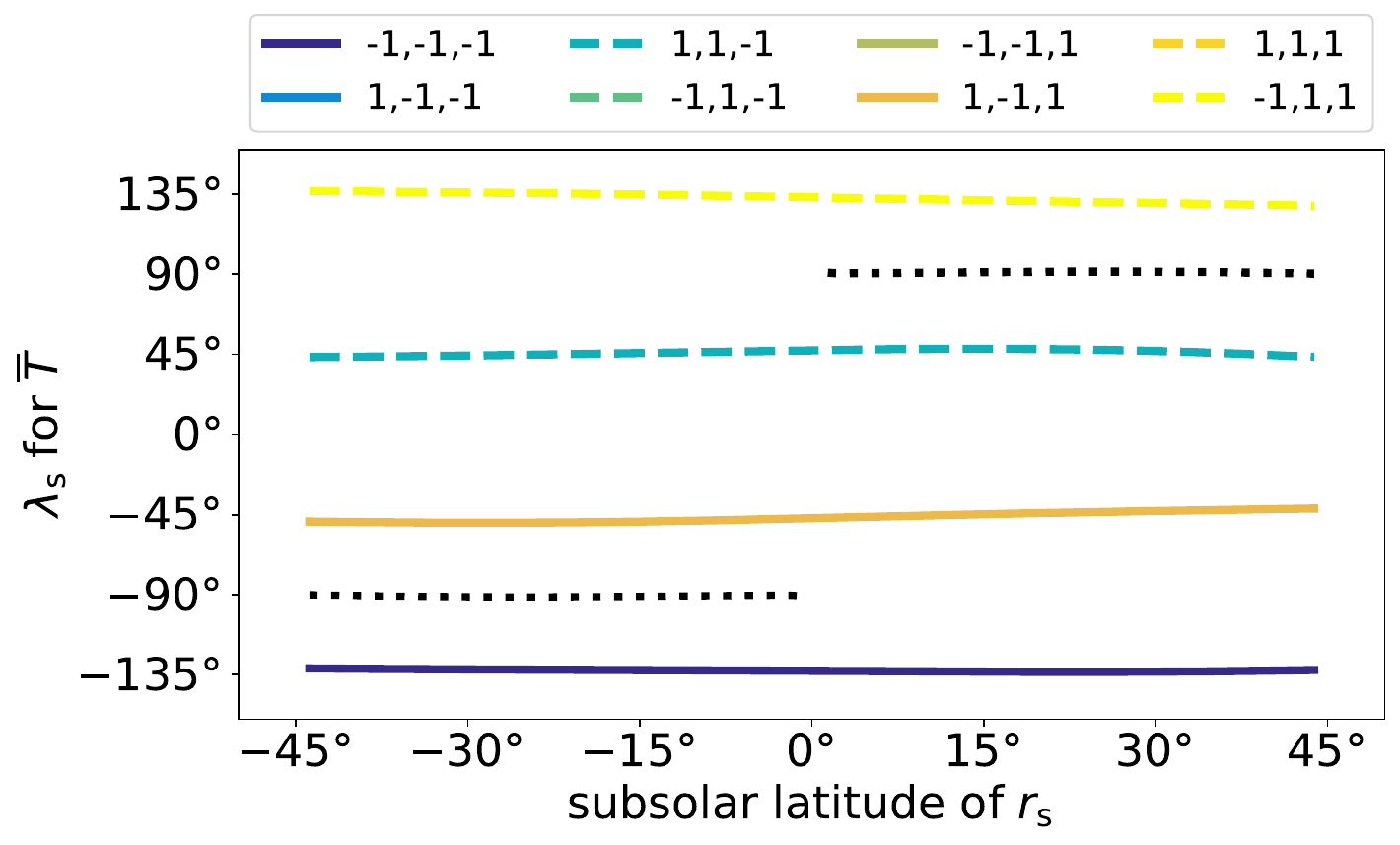}
\caption{Direction of torque, $\overline{T}$, for the near-prolate ellipsoid.
Torque, $\overline{T}$, in Eq.~\eqref{eq:taverage} is shown with model A and
$f_\mathrm{EAF} = \chi_{A_{i,j,k}}$ for region $A_{i,j,k}$ in Eq.~\eqref{eq:regions}.
Uniform activity $f_\mathrm{EAF}=1$ is shown with dotted black lines for the oblate ellipsoid and for the near-prolate ellipsoid.
$\lambda_\mathrm{s}$ and $\phi_\mathrm{s}$ denote longitude and latitude for $\overline{T}$ with respect to the solar basis in Fig.~\ref{fig:solar}.
The solar vector, $r_\mathrm{s}$, changes SSL from $-45^\circ$ to $45^\circ$. 
}
\label{fig:lonoblate}
\end{figure}
From Eq.~\eqref{eq:toinertia} we know that the shape $\Omega$ determines the tensor of inertia $J_\mathrm{BF}$ together with the density function $\rho$.
Complementing this, $\Omega_t$ influences torque formation in Eqs.~\eqref{eq:ngt}, \eqref{eq:taverage}, and \eqref{eq:localt}.
In this section, we assume an instantaneous TPM and a constant solar position, $r_\mathrm{s}$.

We discuss the oblate ellipsoid with uniform activity $f_{\mathrm{EAF}}=1$ on the surface.
The rotational symmetry around $b_3$ yields different properties.
The torque, $T$, is constant in time and finally $T = \overline{T}$.
Both projections $(x\times\nu) \cdot b_{1}(\nu)$ and $(x\times\nu) \cdot b_{3}$ vanish everywhere on $\partial\Omega_t$ with the consequence $T_{1} = T_{3} = 0$.
For southern points $x\in A_\mathrm{SH}$, the feature $0<(x\times\nu) \cdot b_{2}(\nu)$ holds.
In the case of negative SSL ($r_\mathrm{s}\cdot b_3 < 0$), the southern region $A_\mathrm{SH}$ receives more irradiation from the sun so that $v_\mathrm{gas}\dot m$ has higher values than the northern region.
Then the integral in Eq.~\eqref{eq:localt} is dominated by the southern contribution such that $T_2<0$.
The same argument leads to $0<T_2$ for positive SSL.
Using Eq.~\eqref{eq:localt} to evaluate $T\cdot b_{i}(r_\mathrm{s})$  yields $\phi_\mathrm{s} = 0^\circ$ and directions of $T$ that are limited to $\pm b_2(r_\mathrm{s})$.
Fig.~\ref{fig:lonoblate} shows $\lambda_\mathrm{s}$, which flips from $-90^{\circ}$ to $90^{\circ}$ when SSL changes from negative to positive.

The near-prolate ellipsoid is more complex, such that the torque, $T(t)$, is a function of time with uniform activity $f_{\mathrm{EAF}}=1$.
Fig.~\ref{fig:projec} shows the components $(x\times\nu)\cdot b_i(\nu)$ for the near-prolate ellipsoid.
We discuss the plots for the components $b_1$ and $b_3$.
Along circles of constant latitude, $\phi_\mathrm{s}$, for the integrand in Eq.~\eqref{eq:localt}, the distribution is symmetric with respect to zero, such that the integrals are zero, $T_1=T_3=0$, and finally $\overline{T}\cdot b_{1}(r_\mathrm{s}) = \overline{T}\cdot b_{3} = 0$.
$T_2$ in Eq.~\eqref{eq:localt} does not vanish because canceling for $(x\times\nu)\cdot b_2(\nu)$ in Fig.~\ref{fig:projec} does not happen along these line integrals.
This term shows the same signs for points on the hemispheres, $A_\mathrm{SH}$ and $A_\mathrm{NH}$, as discussed for the oblate ellipsoid.
Finally, the directions of $T$ agree for both ellipsoids in Fig.~\ref{fig:lonoblate}.

\begin{figure}[t]
\begin{minipage}{0.48\textwidth}
\centering
\includegraphics[width=0.85\textwidth]{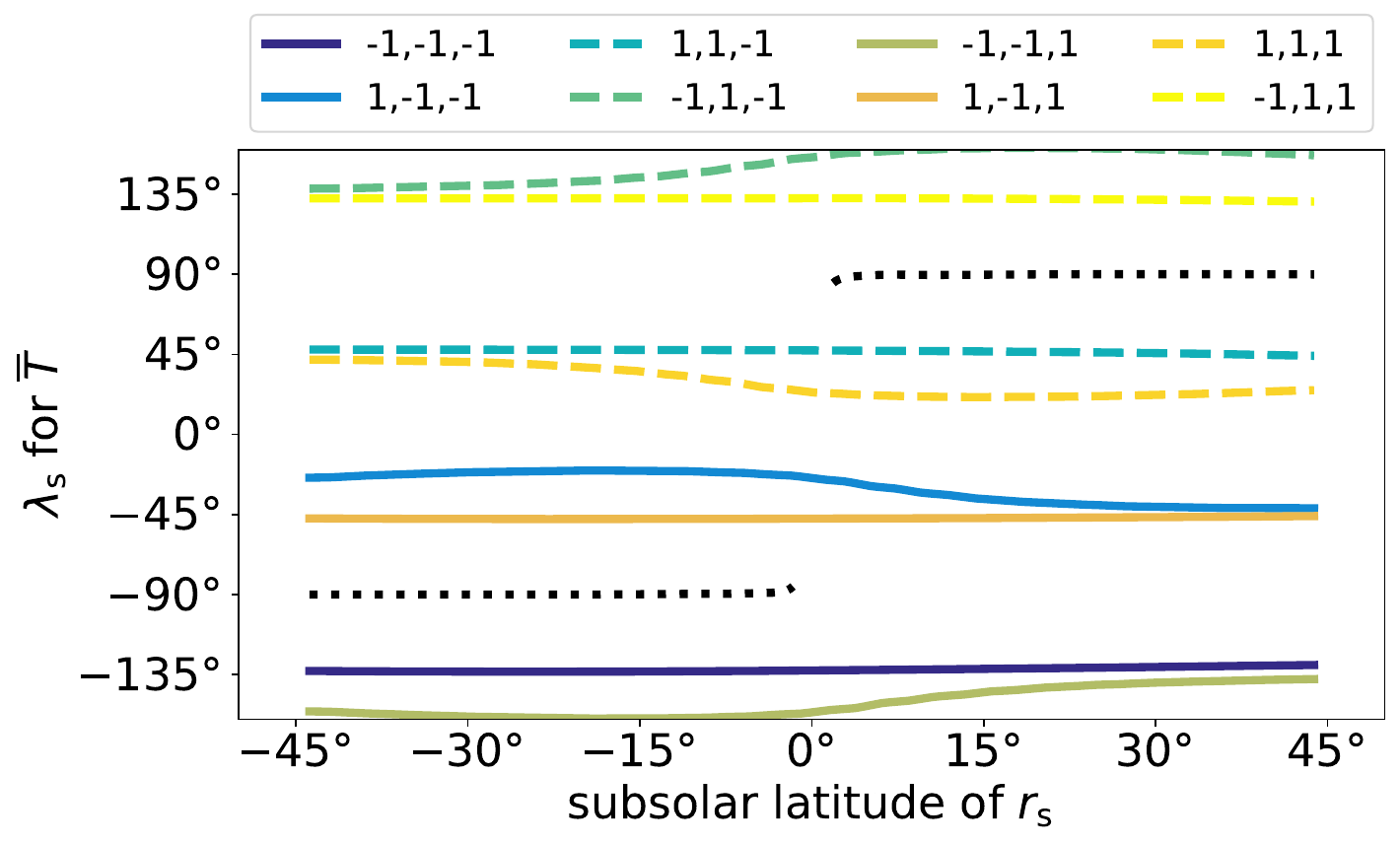}
\end{minipage}
\begin{minipage}{0.48\textwidth}
\centering
\includegraphics[width=0.85\textwidth]{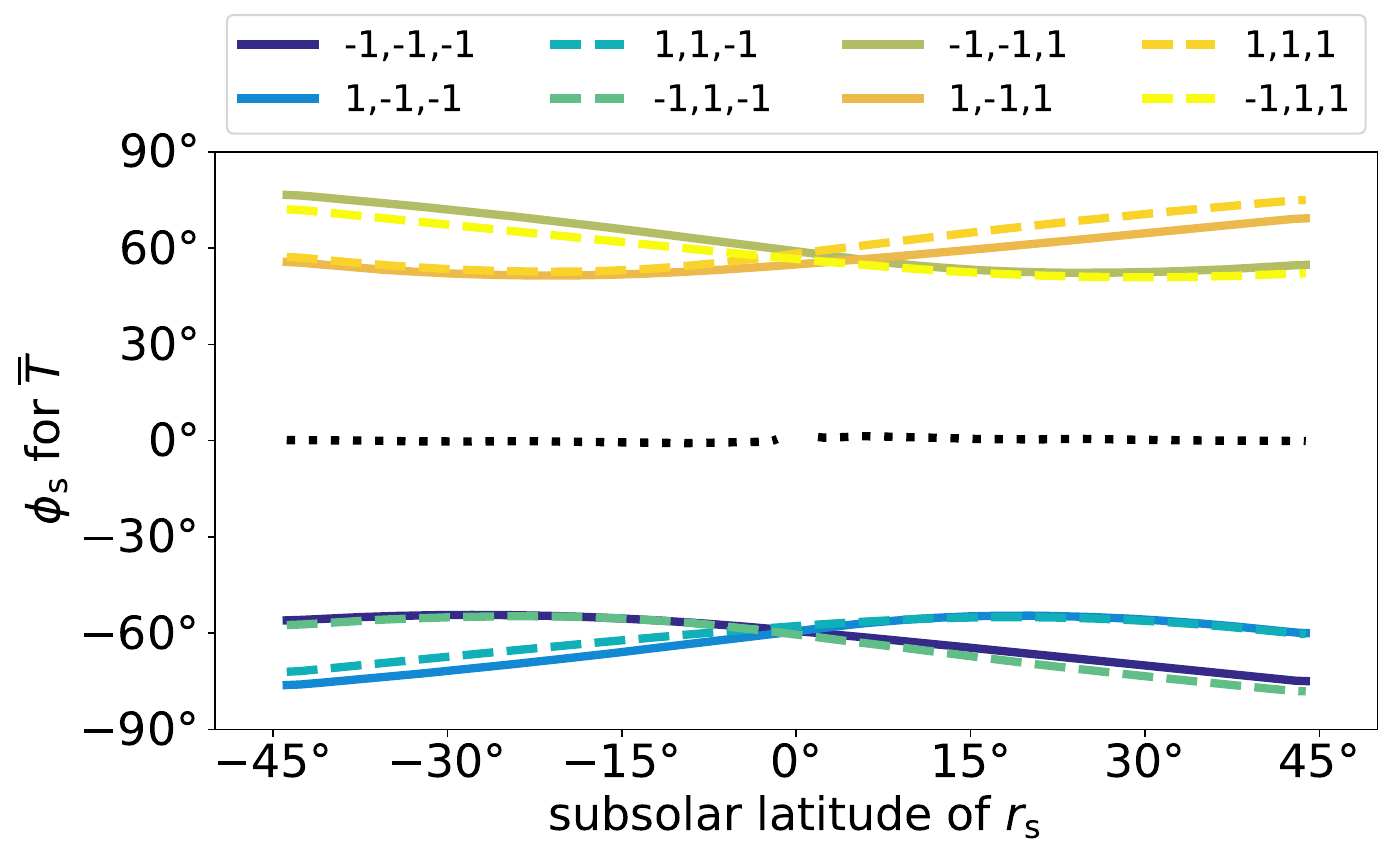}
\end{minipage}
\caption{Direction of torque, $\overline{T}$, for the bilobed shape.
Torque, $\overline{T}$, in Eq.~\eqref{eq:taverage} is shown with model A and
$f_\mathrm{EAF} = \chi_{A_{i,j,k}}$ for region $A_{i,j,k}$ in Eq.~\eqref{eq:regions}.
Uniform activity $f_\mathrm{EAF}=1$ is shown with dotted black lines.
$\lambda_\mathrm{s}$ and $\phi_\mathrm{s}$ denote longitude and latitude for $\overline{T}$ with respect to the solar basis in Fig.~\ref{fig:solar}.
The solar vector, $r_\mathrm{s}$, changes SSL from $-45^\circ$ to $45^\circ$. 
}
\label{fig:lonlatbilobed}
\end{figure}
The properties of the bilobed shape are similar to those of the near-prolate ellipsoid for uniform activity $f_{\mathrm{EAF}}=1$.
$T(t)$ is a function of time.
The two components $\overline{T}\cdot b_{1}(r_\mathrm{s}) = \overline{T}\cdot b_{3} = 0$ vanish.
The corresponding plot of $(x\times\nu)\cdot b_2(\nu)$ for the bilobed shape as in Fig.~\ref{fig:projec} suggests cancelation even for $T_2$.
Shadowing effects must be considered so that the symmetry of $v_\mathrm{gas}\dot m$ is perturbed.
The dotted line in Fig.~\ref{fig:lonlatbilobed} shows the resulting longitude, $\lambda_\mathrm{s}$, of $\overline{T}$.
When comparing the magnitudes of $\overline{T}$, the torque is smaller for the bilobed shape compared to the other shapes.
This goes back to the directional signal caused by shadowing and is more sensitive compared to the geometrical effects of the other shapes.

For all components $T_i$ in Eq.~\eqref{eq:localt} the integrands on the right-hand side have nonzero contributions, which is shown in Fig.~\ref{fig:projec} for the near-prolate ellipsoid (without the term $v_\mathrm{gas}\dot m$).
As a consequence effective active fractions $f_\mathrm{EAF}$ can influence torque formation.
We focus our approach on the superposition of regional contributions in Eq.~\eqref{eq:super}.
$T_i(A_{i,j,k})$ denotes the evaluation of $T_i$ in Eq.~\eqref{eq:localt} assuming the choice $f_\mathrm{EAF} = \chi_{A_{i,j,k}}$.
Because region $A_{i,j,k}$ collects surface points that share the same sign for the integration of $T_l(A_{i,j,k})$, the sign of $T_l$ can be determined with this choice.
The definition of the regions in Eq.~\eqref{eq:regions} yields the properties
\begin{equation}\label{eq:domainconst}
\begin{gathered}
\sgn{}T_1(A_{i,j,k}) = i,\quad
\sgn{}T_2(A_{i,j,k}) = j,\quad
\sgn{}T_3(A_{i,j,k}) = k.
\end{gathered}
\end{equation}
Based on Eq.~\eqref{eq:localt} this extends to the signs of $\overline{T}(A_{i,j,k})\cdot b_{1,2,3}(r_\mathrm{s})$.
For the bilobed shape with model A, Fig.~\ref{fig:lonlatbilobed} shows the longitude, $\lambda_\mathrm{s}$, and latitude, $\phi_\mathrm{s}$, of $\overline{T}(A_{i,j,k})$.
The same figure for the bilobed shape with model $\alpha$ looks similar, leading to the same conclusion.
Eq.~\eqref{eq:domainconst} confirms the assignment of $\lambda_\mathrm{s}$ and $\phi_\mathrm{s}$ to the angular quadrants.
For example $T_1(A_{-1,-1,\pm 1})$ and $T_2(A_{-1,-1,\pm 1})$ both feature negative signs that correspond to the range $-180^\circ < \lambda_\mathrm{s} < -90^\circ$ for the longitude of the vector $\overline{T}(A_{-1,-1,\pm 1})$.
Fig.~\ref{fig:lonlatbilobed} confirms this property.
A closer look at $\lambda_\mathrm{s}$ and $\phi_\mathrm{s}$ reveals that they are close to the middle of the expected ranges.
For the example $A_{-1,-1,\pm 1}$, $\lambda_\mathrm{s}$ is close to $-135^\circ$.
Deviations from the ideal situation go back to the concave shape of this example, leading to self-shadowing, which perturbs the perfect illumination conditions.
The approximation in Eq.~\eqref{eq:localt} is applicable for this case.
For the oblate and near-prolate ellipsoid, this full control of $\lambda_\mathrm{s}$ and $\phi_\mathrm{s}$ for $\overline{T}$ cannot be expected.
For the oblate ellipsoid, all regions $A_{i,j,k}$ vanish and the torque direction is limited to values $\lambda_\mathrm{s} = \pm 90^\circ$.
For the near-prolate ellipsoid in Fig.~\ref{fig:lonoblate} only four of the eight regions are present.

\subsection{Shape of comet 67P/C-G}
\label{sec:67pshape}

\begin{figure}[t]
\begin{minipage}{0.48\textwidth}
\centering
\includegraphics[width=0.85\textwidth]{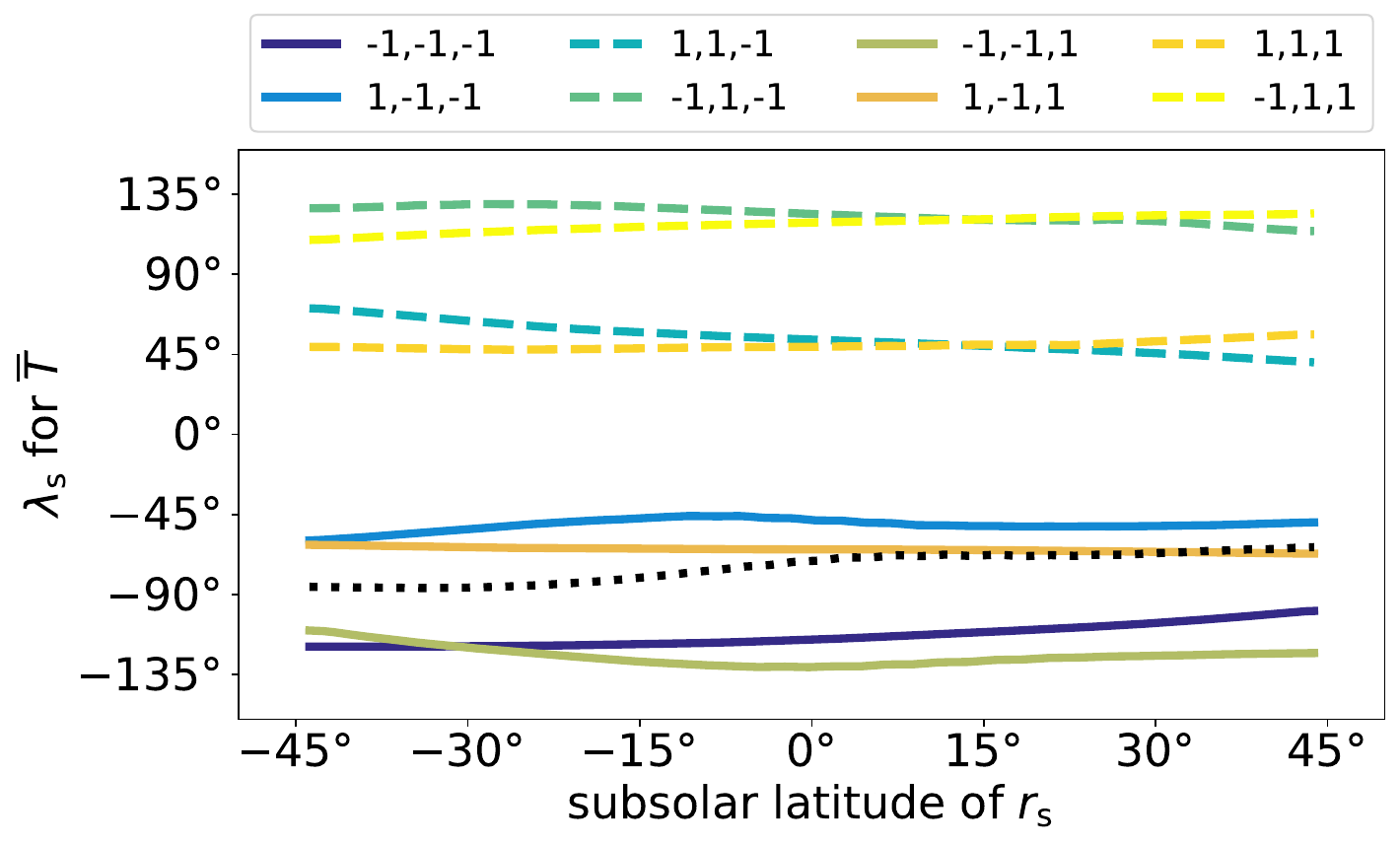}
\end{minipage}
\begin{minipage}{0.48\textwidth}
\centering
\includegraphics[width=0.85\textwidth]{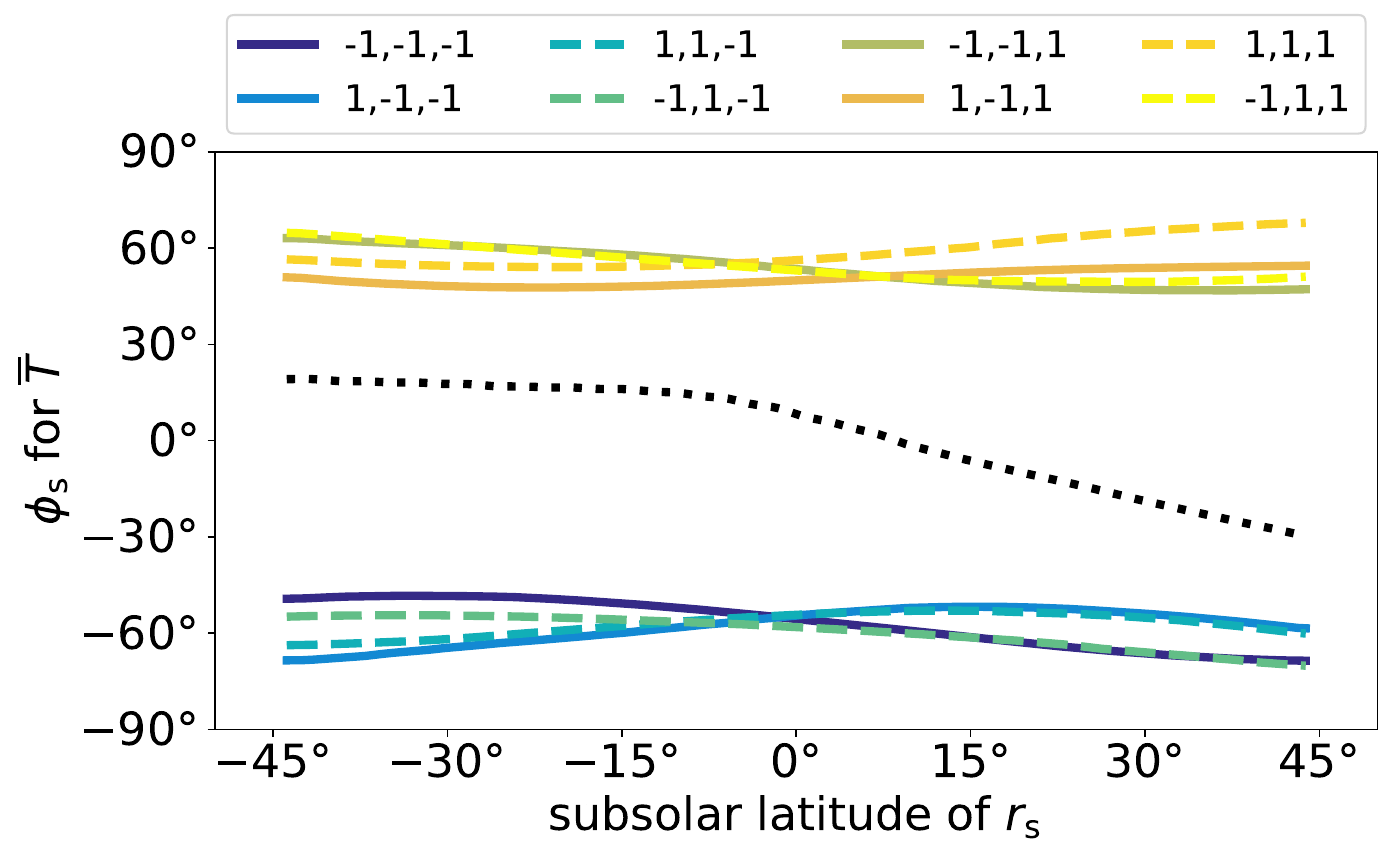}
\end{minipage}
\caption{%
Direction of torque, $\overline{T}$, for the shape 67P.
Torque, $\overline{T}$, in Eq.~\eqref{eq:taverage} is shown with model A and
$f_\mathrm{EAF} = \chi_{A_{i,j,k}}$ for region $A_{i,j,k}$ in Eq.~\eqref{eq:regions}.
Uniform activity $f_\mathrm{EAF}=1$ is shown with dotted black lines.
$\lambda_\mathrm{s}$ and $\phi_\mathrm{s}$ denote longitude and latitude for $\overline{T}$ with respect to the solar basis in Fig.~\ref{fig:solar}.
The solar vector, $r_\mathrm{s}$, changes SSL from $-45^\circ$ to $45^\circ$.
}
\label{fig:lonlat67p}
\end{figure}
The shape of comet 67P is more complex than any of the idealized shapes.
However, the torque formation shares some properties with simplified models, while lack of symmetry brings about new features.
According to the discussion in the last section, we consider an instantaneous TPM.
No component $T_i$ in Eq.~\eqref{eq:localt} is zero and no integrand $(x\times\nu)\cdot b_i(\nu)$ vanishes everywhere on $\partial\Omega_t$.

From Figs.~\ref{fig:lonoblate}, \ref{fig:lonlatbilobed} for both ellipsoids and the bilobed shape, we know that the longitude, $\lambda_\mathrm{s}$, for $\overline{T}$ flips between $-90^\circ$ and $90^\circ$ for changing SSL.
For uniform activity $f_{\mathrm{EAF}}=1$ we investigate this property for shape 67P with Eq.~\eqref{eq:localt}.
Fig.~\ref{fig:projbilobed} shows the component $(x\times\nu)\cdot b_2(\nu)$ on the surface, $\partial\Omega_t$.
Except for the far southern latitudes (small regions around $\phi=-60^\circ$), positive signs are predominant.
For irradiation conditions coming from the north (positive SSL), sublimation controlled by $v_\mathrm{gas}\dot m$ is most active on $A_\mathrm{NH}$ and thus $\overline{T}\cdot b_2(r_\mathrm{s}) \approx T_2 < 0$.
The same argument holds for irradiation coming from mid latitudes even from the south.
For longitude, $\lambda_\mathrm{s}$, of $\overline{T}$ this yields the expectation $-180^\circ < \lambda_\mathrm{s} < 0^\circ$ in a wide range.

For uniform activity, the symmetry of all shapes in the last section leads to vanishing values $\overline{T}\cdot b_3\approx T_3=0$.
This feature is different for shape 67P.
Fig.~\ref{fig:lonlat67p} shows the latitude, $\phi_\mathrm{s}$, for $\overline{T}$ as a falling function with respect to SSL.
In SSL $\approx 10^\circ$, $\phi_\mathrm{s}$ changes the sign from positive to negative and with it $\overline{T}\cdot b_3$.
The southern irradiation conditions (with SSL~$<10^\circ$) related to the predominant sublimation on $A_\mathrm{SH}$ imply $0< \overline{T}\cdot b_3$.
The northern irradiation conditions (with SSL~$>10^\circ$) imply $\overline{T}\cdot b_3 < 0$.
This observation corresponds to an increasing (decreasing) rotation period before (after) the first equinox of the 2015 apparition of comet 67P/C-G as described by \cite{Keller2015a}.
Eq.~\eqref{eq:oneandthree} is a pointwise relation between the components $b_1(\nu)$ and $b_3$ of $x\times\nu$.
This relation does not transfer to the complete integrals $T_1$ and $T_3$ (and not to $\overline{T}\cdot b_1$ and $\overline{T}\cdot b_3$) in Eq.~\eqref{eq:localt}.
We consider this as an indication for $\overline{T}\cdot b_1$ and $\overline{T}\cdot b_3$, only for equally distributed surface activity.
For southern irradiation conditions (and negative SSL) we expect $\sgn \overline{T}\cdot b_1 = \sgn \overline{T}\cdot b_3$.
Due to $0 < \overline{T}\cdot b_3$ we obtain $0 < \overline{T} \cdot b_1(r_\mathrm{s})$ for these conditions.
The same argument for northern irradiation conditions (and positive SSL) yields the same expectation $0 < \overline{T} \cdot b_1(r_\mathrm{s})$.
For longitude, $\lambda_\mathrm{s}$, of $\overline{T}$ this gives the constraint $-90^\circ < \lambda_\mathrm{s} < 90^\circ$.
Together with the analysis of the $b_2$ component in the last paragraph, we obtain the expectation $-90^\circ < \lambda_\mathrm{s} < 0^\circ$.
Fig.~\ref{fig:lonlat67p} confirms this, showing that the course of $\lambda_\mathrm{s}$ is in the range between $-90^\circ$ and $-70^\circ$.

For shape 67P the eight regions $A_{i,j,k}$ defined in Eq.~\eqref{eq:regions} are shown in the Figs.~\ref{fig:regbilobed}, \ref{fig:reg3d}.
The orientations of $\overline{T}(A_{i,j,k})$ are shown in Fig.~\ref{fig:lonlat67p} based on effective active fractions $f_{\mathrm{EAF}} = \chi_{A_{i,j,k}}$.
The longitude, $\lambda_\mathrm{s}$, and latitude, $\phi_\mathrm{s}$, of the different $\overline{T}$ are within the ranges $-135^\circ< \lambda_\mathrm{s} < 135^\circ$ and $-75^\circ < \phi_\mathrm{s} < 75^\circ$.
This is similar to Fig.~\ref{fig:lonlatbilobed} for the bilobed shape.
The appropriate weights for the superposition in Eq.~\eqref{eq:super} will be able to create torque formation for the prescribed values of $\lambda_\mathrm{s}$ and $\phi_\mathrm{s}$.
Because the argumentation in Sect.~\ref{sec:simpleshape} is independent of the specific shape, Eq.~\eqref{eq:domainconst} is true for shape 67P as well.

\section{Rotational angular momentum and velocity}
\label{sec:velocity}

We studied the evolution of the angular velocity, $\omega$.
Similarly to the discussion for $T$, the vector $\omega$ can be decomposed into components with respect to the solar basis, $B(r_\mathrm{s})$, in Fig.~\ref{fig:solar}.
All initial states within the present section, $R_0$ and $L_0$, are assumed to be in lowest energy.
In this case, the axes $b_3$, $\omega$, and $L$ (principal axis, rotation axis, and rotational angular momentum) are aligned to each other.
The presence of torque, $T$, causes an excitation for the rotation state such that $b_3$, $\omega$, and $L$ are no longer parallel.
For rotation states close to lowest energy, the torque formation discussed in Sect.~\ref{sec:torque} remains valid qualitatively.
The longitude, $\lambda_\mathrm{s}$, and the latitude, $\phi_\mathrm{s}$, of $\omega$ with respect to the solar basis are one option to express the alignment between $\omega$ and $b_3$.
For the oblate ellipsoid, the near-prolate ellipsoid, the bilobed shape and the shape 67P, we considered model A with $f_\mathrm{EAF}=1/8$.
For the shape 67P, we considered a second variant based on the model $\alpha$ with $f_\mathrm{EAF}=1/4$.

To study the evolution of the rotation state, we studied the formation of the torque, $T$, according to Eq.~\ref{eq:ngt}.
The first setup considers a prescribed vector for the solar direction $r_\mathrm{s}$, a constant heliocentric distance $|r_\mathrm{s}| = 2\, \mathrm{au}$ and a changing SSL from $-45^\circ$ to $45^\circ$ during a time span of 800~days.
For this setup, the latitude, $\phi_\mathrm{s}$, shows small deviations in the range $89.99^\circ \leq  \phi_\mathrm{s} \leq 90^\circ$.
This significantly improves the analysis of the linearized equations ($\gamma\approx 0.5^\circ$) in Sect.~\ref{sec:systems} and shows the close alignment of the vectors $\omega$ and $b_3$.
The rotation remains close to a lowest-energy state.
At the same time, the longitude, $\lambda_\mathrm{s}$, for $\omega$ covers the entire angular range $-180^\circ \leq \lambda_\mathrm{s} \leq 180^\circ$ that describes the precessional movement of $\omega$ surrounding $b_3$.

\subsection{Uniform activity}

\begin{figure}[t]
\centering
\includegraphics[width=0.3\textwidth]{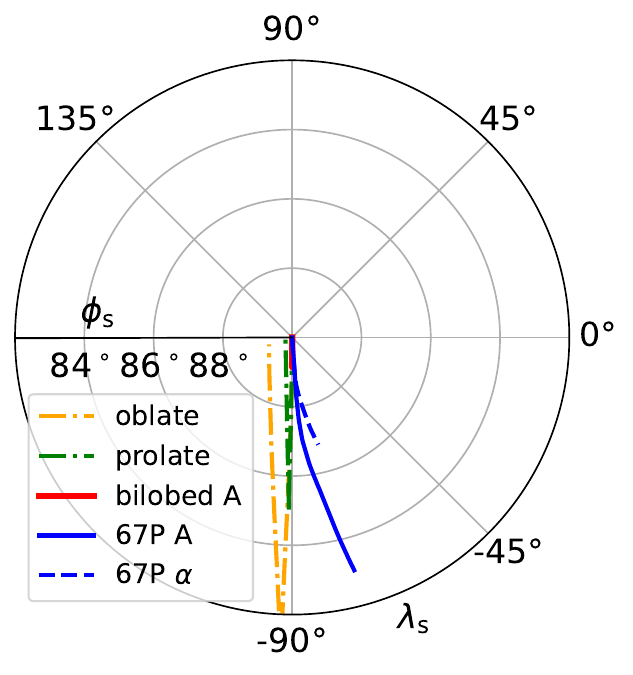}
\caption{Longitude, $\lambda_\mathrm{s}$, and latitude, $\phi_\mathrm{s}$, for angular velocity, $\omega$, with respect to the fixed solar basis in Fig.~\ref{fig:solar}, based on $r_\mathrm{s}(t_0)$ and $b_3 = R(t_0)e_3$ at initial time, $t_0$.
$T$ is evaluated with $|r_\mathrm{s}| = 2\,\mathrm{au}$ and changing SSL from $-45^\circ$ to $45^\circ$.
Considered are the oblate ellipsoid, the near-prolate ellipsoid, the bilobed shape, the shape 67P with model A/$f_\mathrm{EAF}=1/8$, and the shape 67P with model $\alpha$/$f_\mathrm{EAF}=1/4$.
}\label{fig:polarngt}
\end{figure}
We consider a uniformly distributed surface ice coverage.
In this case, the principal axis, $b_3 = R(t) e_3$, changes in time.
To plot the evolution of $\omega$ in figures with respect to an inertial frame, we fix the solar basis of Fig.~\ref{fig:solar} based on the state at initial time $t_0$, namely with $r_\mathrm{s}(t_0)$ and $b_3 = R(t_0) e_3$.
Fig.~\ref{fig:polarngt} shows $\lambda_\mathrm{s}$ and $\phi_\mathrm{s}$ for $\omega$ with respect to this fixed solar basis.
Due to lowest-energy state at time $t_0$, $\omega$ and $b_3$ start aligned, which is reflected by $\phi_\mathrm{s} = 90^\circ$.
The close alignment for $b_3$, $\omega$, and $L$ implies that Fig.~\ref{fig:polarngt} shows the directions for the principal axis, $b_3(t)$, and for rotational angular momentum, $L(t)$, up to the order of $10^{-2}$ degree.
The torques in Sect.~\ref{sec:torque} cause directional changes on the order of $5^\circ$ for $b_3$, $\omega$, and $L$.
Due to the smaller torque values for the bilobed shape, the changes in $\omega$ are smaller by an order.
Qualitatively, the shapes evolve $\omega$ along the line $\lambda_\mathrm{s} = -90^\circ$.
According to Sect.~\ref{sec:systems}, the equatorial projections of $r_\mathrm{s}$ and $\omega$ are close to perpendicular.
The torque orientations in Figs.~\ref{fig:lonoblate}, \ref{fig:lonlat67p} confirm important properties of the evolution of $\omega$.
At SSL~$=0^\circ$ for both ellipsoids the jump of torque longitude from $-90^\circ$ to $90^\circ$  is reflected by corresponding turning points for the $\omega$ trajectory in Fig.~\ref{fig:polarngt}.
For shape 67P the $\omega$ trajectories (model A and model $\alpha$) do not contain a turning point.
This observation agrees with the smooth curve for longitude, $\lambda_\mathrm{s}$, in Fig.~\ref{fig:lonlat67p}. 
Longitude, $\lambda_\mathrm{s}$, is in the range between $-90^\circ$ and $-70^\circ$, which explains the deviation from $\lambda_\mathrm{s} = -90^\circ$ for $\omega$ in Fig.~\ref{fig:polarngt}.

\begin{figure}[t]
\begin{minipage}{0.48\textwidth}
\centering
\includegraphics[width=0.95\textwidth]{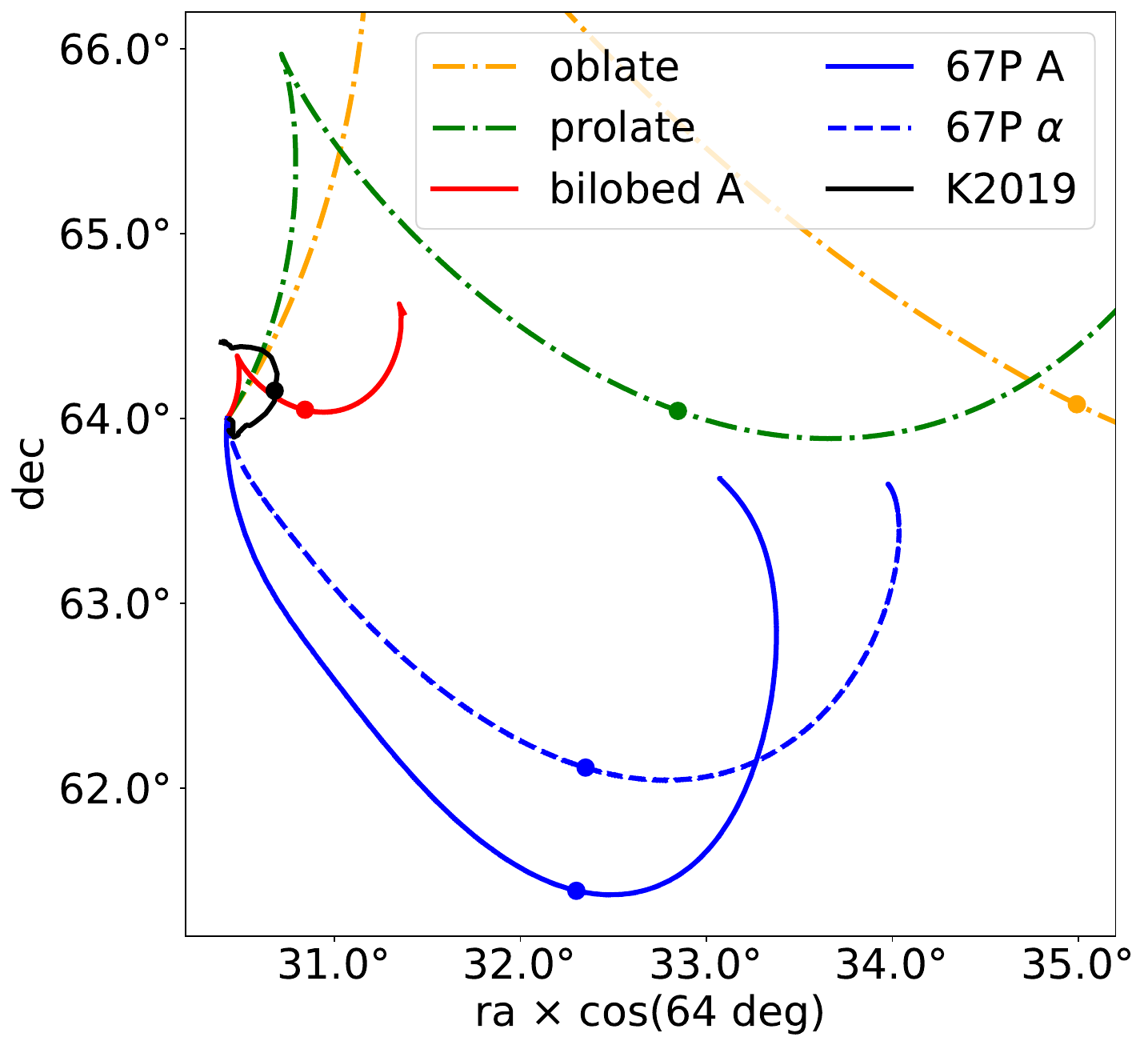}
\end{minipage}
\begin{minipage}{0.48\textwidth}
\centering
\includegraphics[width=0.95\textwidth]{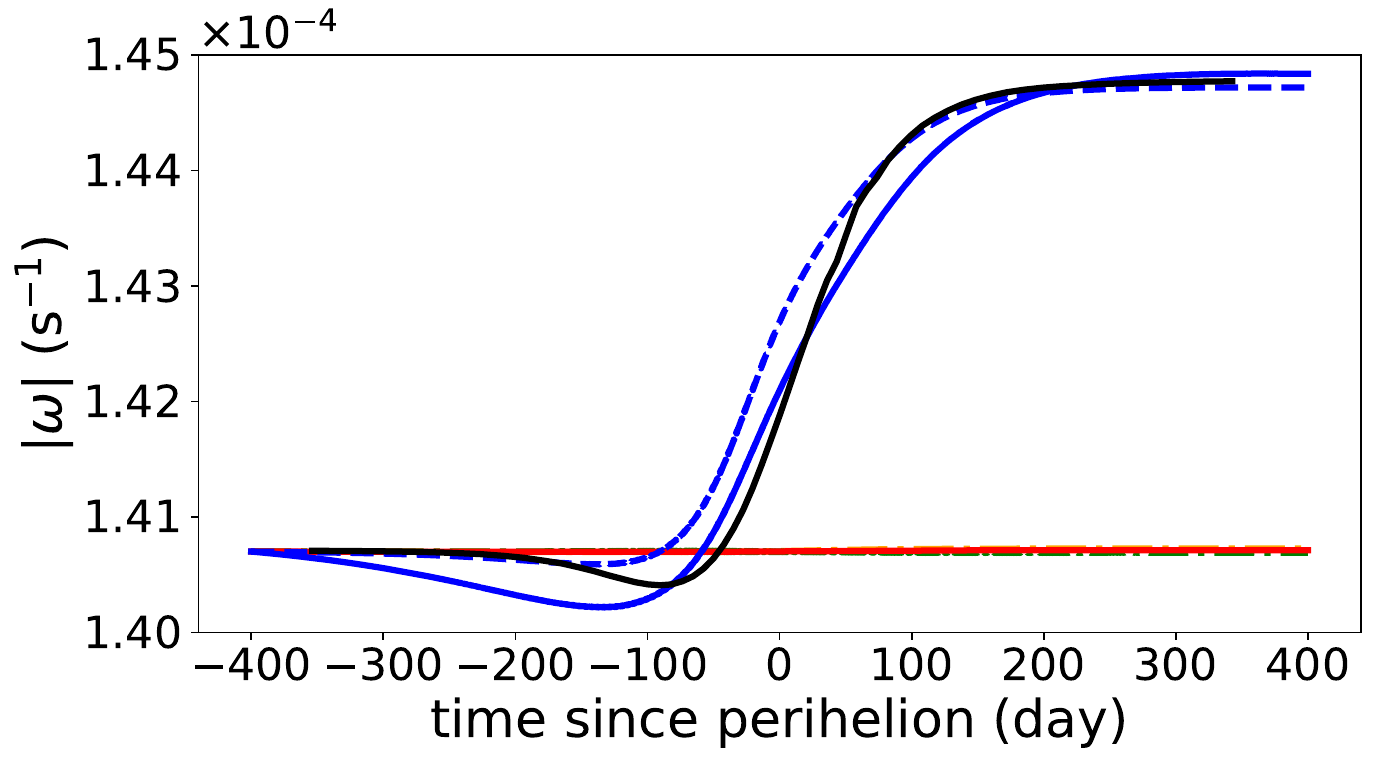}
\end{minipage}
\caption{Angular velocity, $\omega$.
$R_0$, $L_0$ is in lowest-energy state.
$T$ is evaluated for solar vector, $r_\mathrm{s}$, describing the trajectory of the sun during the apparition 2015 relative to comet 67P/C-G.
Considered are the oblate ellipsoid, the near-prolate ellipsoid, the bilobed shape, the shape 67P with model A/$f_\mathrm{EAF}=1/8$, and the shape 67P with model $\alpha$/$f_\mathrm{EAF}=1/4$.
The markers on each curve denote the $\omega$ direction at perihelion passage in August 2015.
Top: Right ascension (ra) and declination (dec) for $\omega$ in the equatorial J2000 (EME2000) reference frame.
Bottom: Angular frequency, $|\omega|$.
}
\label{fig:app2015}
\end{figure}
Fig.~\ref{fig:app2015} shows the evolution of $\omega$ in time from 400 days before to 400 days after the perihelion passage of comet 67P/C-G.
At initial time, the evolution starts with a fixed angular velocity, $\omega$, and a lowest-energy state.
Unlike Fig.~\ref{fig:polarngt}, the solar vector, $r_\mathrm{s}$, describes the trajectory of the sun during apparition 2015 relative to comet 67P/C-G.
In the beginning, before the first equinox, SSL is positive and irradiation comes from the north.
For both ellipsoids in Fig.~\ref{fig:lonoblate}, we know that the longitude, $\lambda_\mathrm{s}$, for $\overline{T}$ is at $90^\circ$.
For shape 67P in Fig.~\ref{fig:lonlat67p}, $\lambda_\mathrm{s}$ for $\overline{T}$ is close to $-75^\circ$.
In Fig.~\ref{fig:app2015} these opposite directions are reflected in the $\omega$ trajectories.
At the two equinox times (95 days before and 224 days after perihelion), the SSL is zero.
For the two ellipsoids, these points are related to directional switches between $-90^\circ$ and $90^\circ$ for $\lambda_\mathrm{s}$ of $\overline{T}$ in Fig.~\ref{fig:lonoblate}.
For the corresponding $\omega$ trajectories this is related to turning points.
This geometric phenomenon is known and reported; for example, from \cite{Whipple1979}, \cite{Sekanina1984}, and \cite{Gutierrez2007}.
Between the two equinox times, the SSL is negative.
This setup, shown in Figs.~\ref{fig:lonoblate}, \ref{fig:lonlat67p}, leads to torque directions close to $\lambda_\mathrm{s} = -90^\circ$ for all shapes.
Fig.~\ref{fig:app2015} reflects this by showing similar slopes along all curves around the perihelion passage (marked by a point on the curve).
For the bilobed shape, the relative small magnitude of the torque (discussed in Sect.~\ref{sec:torque}) is represented by the small movement of $\omega$.
For both ellipsoids and the bilobed shape, the latitude, $\phi_\mathrm{s}$, of $\overline{T}$ is always zero, which yields a constant rotation period, as shown in Fig.~\ref{fig:app2015}.
For the shape 67P, the decrease in $|\omega|$ until day -80 (followed by an increase) is directly related to the positive SSL and the corresponding torque direction (with $\phi_\mathrm{s}<0$) in Fig.~\ref{fig:lonlat67p}.

\subsection{Weighted activity}

In Fig.~\ref{fig:app2015} the values $f_\mathrm{EAF}=1/8$ for modal A and $f_\mathrm{EAF}=1/4$ for model $\alpha$ are chosen such that the angular frequency, $|\omega|$, changes by the same order of magnitude as the observational data for comet 67P.
The corresponding modeled changes for the $\omega$ direction (ra and dec) are on the order of $2^\circ-3^\circ$, which is too high by the order of four to six compared to the observation.
This property has been reported by \cite{Kramer2019a} and \cite{Attree2024}.
The second observation for the shape 67P in Fig.~\ref{fig:app2015} applies to the slope along the $\omega$ trajectory at perihelion (and during the complete pathway).
Those curves seem to be rotated between models and observation.
A uniform activity on the surface of the shape 67P with either models A or $\alpha$ is not able to bring into agreement the model results and the observations in 2015.
This limitation can be resolved by the weighted activity in Eq.~\eqref{eq:super}.
With the assumption of a lowest-energy state and a fixed phase angle for $R_0$ at initial time, an initial vector $\omega_0\in\real^3$ for angular velocity (three parameters) determines the complete initial condition, $R_0$ and $L_0$, for the solution of Eq.~\eqref{eq:rotation}.
Because the activity weights $f_{\mathrm{EAF},i,j,k}$ in Eq.~\eqref{eq:super} are eight parameters, a model evolution has $n_\mathrm{P}=11$ open parameters $p_i$ for $i=1,...,n_\mathrm{P}$.
We fitted the model to the data in Eq.~\eqref{eq:data} and minimized the quadratic norm,
\begin{gather}\label{eq:sigma}
\sigma^2
= \sum_{i=1}^{N_\mathrm{time}} \sum_{j=1}^{3}
\left( \frac{\data_j(\omega(t_i)) - data_{i,j}}{\sigma_{i,j}} \right)^2.
\end{gather}
Because the mapping from the parameter vector $( p_1,..., p_{n_\mathrm{P}})$ to the vector $( \omega(t_1),..., \omega(t_{N_\mathrm{time}}) )$ is nonlinear, we solved the minimization problem for $\sigma$ with a standard Gauss-Newton method.
The solver attributes negative values to $f_{\mathrm{EAF},-1,1,-1}$ and $f_{\mathrm{EAF},-1,1,1}$, so that we set both values to zero.
These locations do not agree with low activity of \cite{Kramer2019a}.
This agreement cannot be expected because the patch definitions of \cite{Kramer2019a} mix contributions of several regions $A_{i,j,k}$ from Eq.~\eqref{eq:regions} and optimize the rotation axis under the additional constraint to keep the overall active surface fraction as homogeneous as possible.

\begin{figure}[t]
\centering
\includegraphics[width=0.48\textwidth]{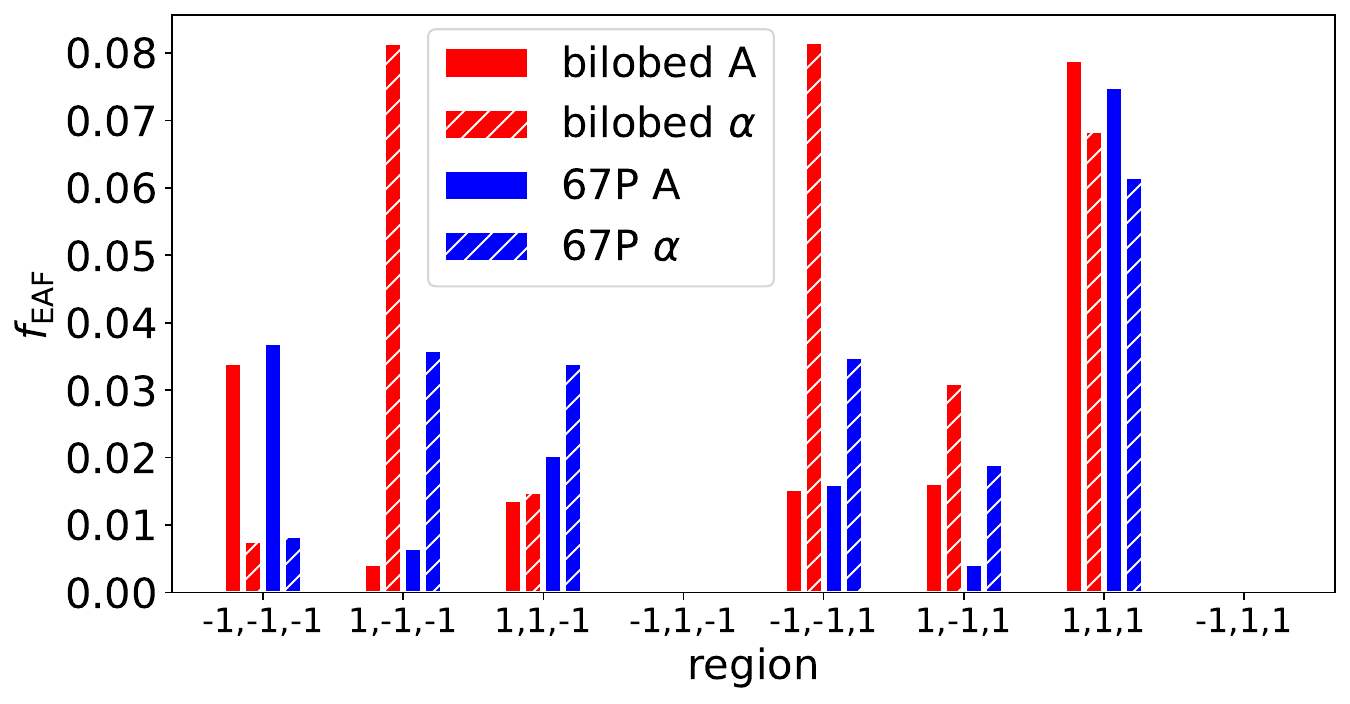}
\caption{Activity weights $f_{\mathrm{EAF},i,j,k}$ for regions $A_{i,j,k}$ in Eq.~\eqref{eq:super} for the bilobed shape and the shape 67P with model A and model $\alpha$.
The values of $f_{\mathrm{EAF},i,j,k}$ are constrained by the $\omega$ data in Sect.~\ref{sec:data}.
}
\label{fig:eaf}
\end{figure}
\begin{figure}[t]
\begin{minipage}{0.48\textwidth}
\centering
\includegraphics[width=0.95\textwidth]{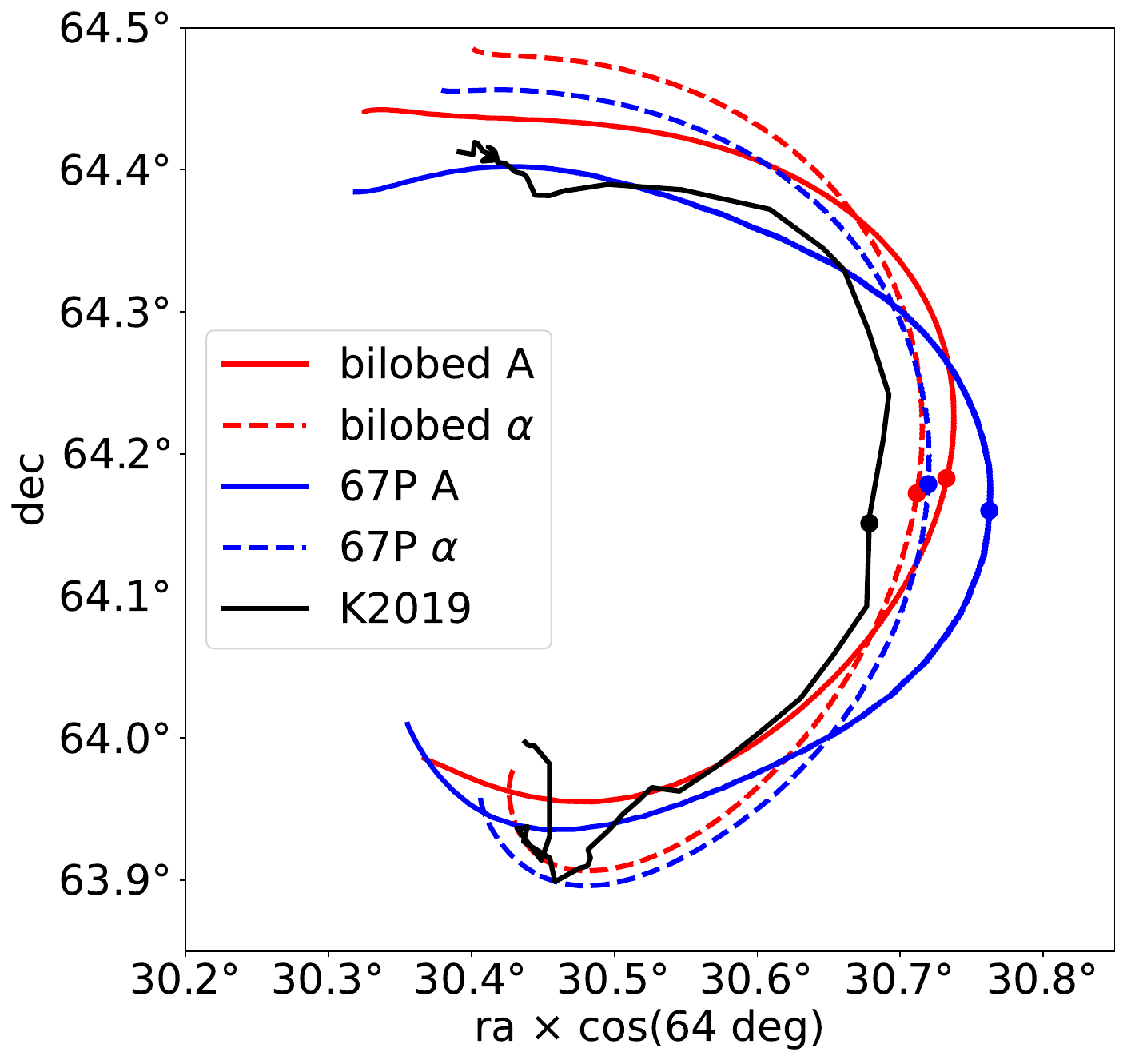}
\end{minipage}
\begin{minipage}{0.48\textwidth}
\centering
\includegraphics[width=0.95\textwidth]{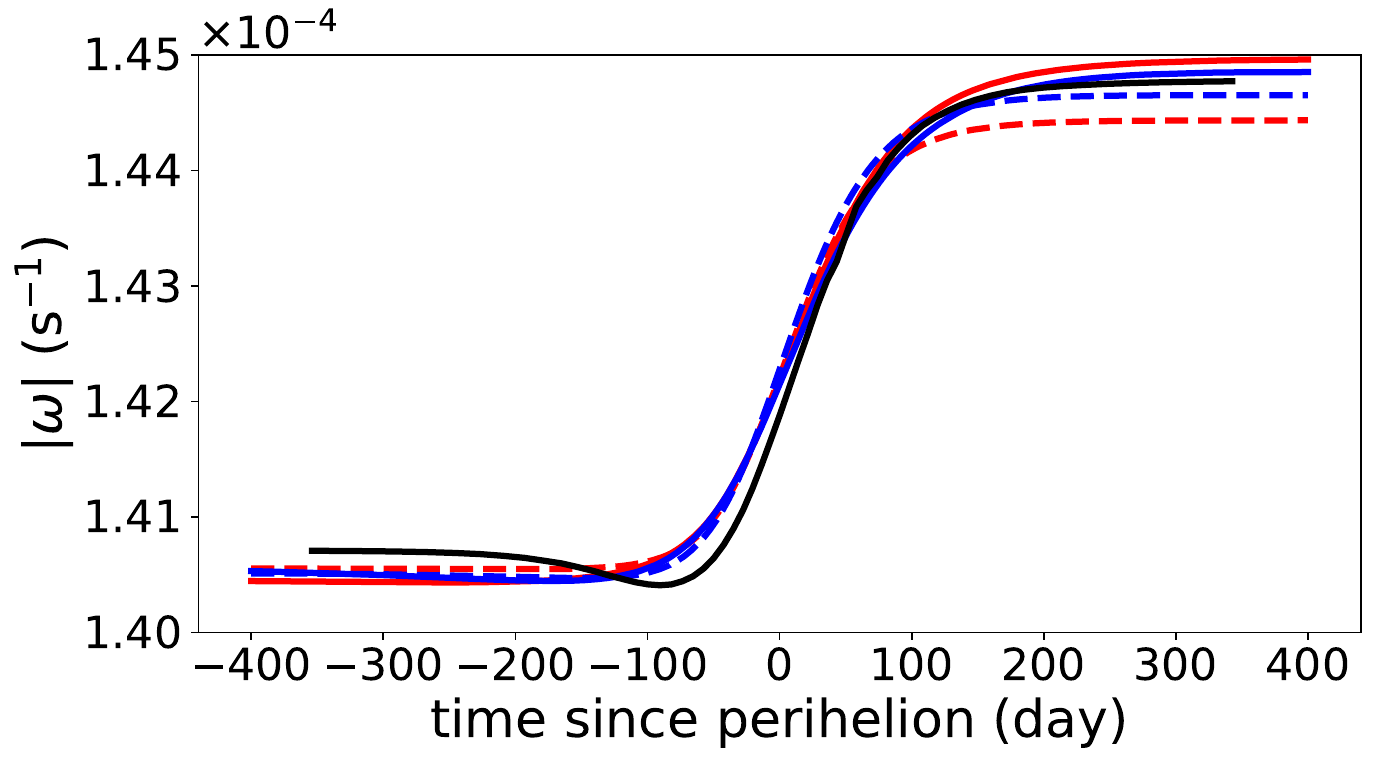}
\end{minipage}
\caption{Angular velocity, $\omega$, similar to Fig.~\ref{fig:app2015} for
the bilobed shape and the shape 67P with model A and model $\alpha$.
Weighted activity from Fig.~\ref{fig:eaf}.
Top: Right ascension (ra) and declination (dec) for $\omega$.
Bottom: Angular frequency, $|\omega|$.
}
\label{fig:data01}
\end{figure}
We considered four setups with different solutions for the weighted surface activity.
These were the bilobed shape and the shape 67P, each in combination with model A and model $\alpha$.
The two ellipsoids were not considered, because they do not have eight nonempty regions $A_{i,j,k}$.
Fig.~\ref{fig:eaf} shows the resulting activity weights $f_{\mathrm{EAF},i,j,k}$.
Fig.~\ref{fig:data01} shows the associated evolution for $\omega$.
The norms $\sigma$ in Eq.~\eqref{eq:sigma} are as good as $0.1^\circ$ for the directional fraction (ra and dec) and as $75\,\mathrm{s}$ for the fraction describing the rotation period.
Modifications to model configurations produce fits with similar accuracy, $\sigma$, and allow us to draw the same conclusions.
The first kind of modification is to change the number of elements and, along with it, the shape resolution that we discussed in Sect.~\ref{sec:systems}.
For the bilobed shape half and double the number of elements $N_\mathrm{E}$ change the value of $\sigma$ in the range of up to $0.01\sigma$.
The second kind refers to the bilobed shape with the modified value $\alpha=1.5$ that we introduced in Sect.~\ref{sec:tpm}.
Similarly to the variations in Fig.~\ref{fig:eaf}, an agreement on the order of 25\% is obtained only for $f_{\mathrm{EAF},1,1,1}$.
For all remaining regions, the uncertainty is on the order of two or greater.
The last kind of modification refers to variations in the sampling rate and uncertainty weights for the data set in Sect.~\ref{sec:data}.
A variation in the data incorporating an equidistant time sampling of 7\,days and constant uncertainties, $\sigma_i$, of $0.05^\circ$ for direction and 37\,s for time period does not change the results.

For the four model setups, the solutions in Fig.~\ref{fig:data01} solve the mismatch between models and observations in Fig.~\ref{fig:app2015}.
The general changes for the $\omega$ direction and for the angular frequency, $|\omega|$, are qualitatively consistent with the data in Sect.~\ref{sec:data}.
For all setups, the weights $f_{\mathrm{EAF},i,j,k}$ in Fig.~\ref{fig:eaf} are significantly smaller than for uniform activity, on the order of three for model A and on the order of five for model $\alpha$.
Because $f_{\mathrm{EAF}}$ is a unit free quantity, different values $f_{\mathrm{EAF},i,j,k}$ for model A and model $\alpha$ go back to the normalization constants for both TPMs in Sect.~\ref{sec:tpm}.
All values of the effective active fraction are bounded by $f_{\mathrm{EAF}} \leq 0.085$.
In the regions $A_{-1,-1,-1}, A_{1,\pm 1,1} \subset A_\mathrm{SH}$ and $A_{1,\pm 1,-1}, A_{-1,-1,1} \subset A_\mathrm{NH}$ all setups agree to attribute nonzero activity.
The uncertainties for the weights $f_{\mathrm{EAF},i,j,k}$ suffer from significant limitations.
The only region in which all the setups agree is $A_{1,1,1}\subset A_\mathrm{SH}$ with weights in the range $0.06 \leq f_{\mathrm{EAF},1,1,1} \leq 0.08$.
This is related to the fact that the data in Sect.~\ref{sec:data} describe the dynamics for negative SSL (during the perihelion passage of comet 67P/C-G in 2015) predominantly and thus for activity on $A_\mathrm{SH}$.
In Fig.~\ref{fig:regbilobed} for the shape 67P, $A_{1,1,1}$ is located in the regions $(0^\circ, 60^\circ)\times (-90^\circ,-45^\circ)$ and $(150^\circ, 210^\circ)\times (-90^\circ,-30^\circ)$ for $\lambda$ and $\phi$.
This fits well with the location of increased activity in Fig.~10 of \cite{Kramer2019a}.
The same authors report a decrease in the effective active fraction in the regions $(-45^\circ, 15^\circ)\times (-30^\circ,0^\circ)$ and $(150^\circ, 210^\circ)\times (-15^\circ,15^\circ)$ for $\lambda$ and $\phi$.
This statement correlates to reduced weights in Fig.~\ref{fig:eaf} for the regions $A_{-1,-1,-1}$ and $A_{1,-1,1}$ in Fig.~\ref{fig:regbilobed}.
Restricted to model A, some agreement can be obtained between the two shapes on the regions $A_{-1,-1,-1}$, $A_{1,1,-1}$, and $A_{-1,-1,1}$.
On the remaining regions the weights contain much higher uncertainties based on the consideration of all different setups.

\begin{figure}[t]
\centering
\includegraphics[width=0.35\textwidth]{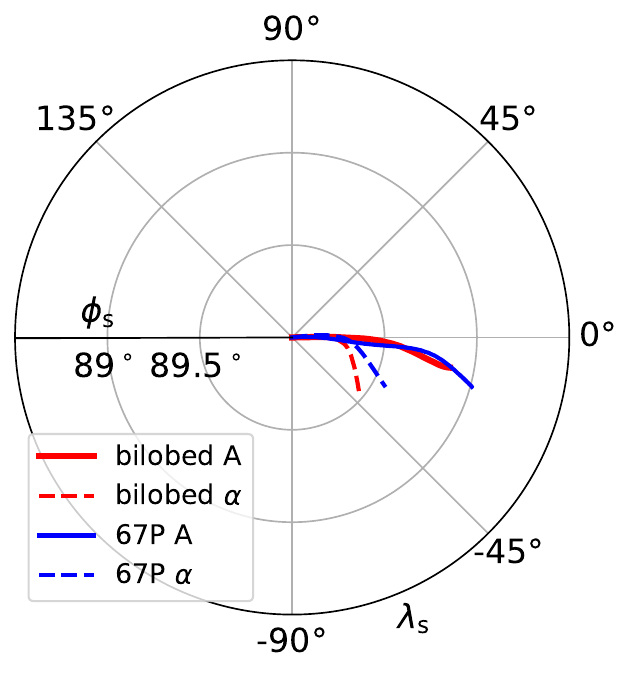}
\caption{Longitude, $\lambda_\mathrm{s}$, and latitude, $\phi_\mathrm{s}$, for angular velocity, $\omega$. The same projection with respect to the fixed solar basis and the same $r_\mathrm{s}$ as in Fig.~\ref{fig:polarngt}.
Considered are the bilobed shape with model A, the shape 67P with model A, and the shape 67P with model $\alpha$, and weighted activity $f_\mathrm{EAF}$ in Fig.~\ref{fig:eaf}.
}
\label{fig:polaropt}
\end{figure}
One major difference between the evolutions of uniform (Fig.~\ref{fig:app2015}) and weighted 
(Fig.~\ref{fig:data01}) activity is the rotated trajectory for the two $\omega$ directions.
For uniform activity, Fig.~\ref{fig:polarngt} shows this trajectory (with respect to the solar basis) close to the line $\lambda_\mathrm{s} = -90^\circ$.
For weighted activity, Fig.~\ref{fig:polaropt} shows the same trajectory close to the line $\lambda_\mathrm{s} = 0^\circ$.
Especially for negative SSLs (related to the early part of the evolution in Fig.~\ref{fig:polaropt}), weighted activity causes this rotation for the $\omega$ trajectory (and for the slope along the trajectory) by approximately $90^\circ$.
Negative SSLs represent predominant illumination conditions during the perihelion passage of comet 67P/C-G in 2015.
Due to rotation states close to lowest energy, this directional change applies to $L$ as well.
For weighted activity, the torque, $\overline{T}$, points mainly in the $\lambda_\mathrm{s} = 0^\circ$ direction that expresses $\overline{T}\cdot b_1 > 0$.
Thus this directional evolution in Fig.~\ref{fig:polaropt} confirms the increased weights for region $A_{1,1,1}$ in Fig.~\ref{fig:eaf}.

\section{Conclusions}

The vectorial torque efficiency introduced in Eq.~\eqref{eq:vte} can be evaluated for any general cometary body.
It provides a generalization of the scalar torque efficiency in \cite{Keller2015} and \cite{Attree2019} that controls the rotational period alone.
The signs of all three components of the vectorial torque efficiency lead to the specification of a domain decomposition into a maximum of eight distinct regions on the surface.
In particular, this applies to the bilobed shape as well as to the body shape of comet 67P/C-G.
Geometric symmetry for idealized cases can reduce the number of regions.
The oblate ellipsoid features zero and the near-prolate ellipsoid features four nonempty regions.
The domain decomposition is divided into two subsets containing four regions each.
One subset is linked to the southern hemisphere, $A_\mathrm{SH}$, and the other is linked to the northern hemisphere, $A_\mathrm{NH}$.

For comet 67P/C-G in the time between 2014 and 2016, uniform activity leads to a mismatch between model and observations for directions of the rotation state.
This mismatch is resolved with a best-fit solution for effective active fractions.
This directional switch by approximately $90^\circ$ is illustrated in Fig.~\ref{fig:data01} for angular velocity and in Fig.~\ref{fig:polaropt} for torque. 
The fit quality for the data is on the order of $0.1^\circ$ for the $\omega$ direction and $75~\mathrm{s}$ for the rotation period.
Only one region $A_{1,1,1}$ is constrained by the data with values at $0.06\leq f_{\mathrm{EAF},1,1,1} \leq 0.08$.
It is located on the southern hemisphere, $A_\mathrm{SH}$, which is the hemisphere that has a high solar irradiation (and negative SSL) around the perihelion passage of comet 67P/C-G in 2015.
Thus, torque activity at that time is high for $A_{1,1,1}$ and according to the construction of $A_{1,1,1}$ the torque direction has three positive components with respect to the solar basis.
In particular, this produces a positive $b_3$ component for the torque, an increase in the angular frequency, $|\omega|$, and a decrease in the rotation period at that time.
For the weights $f_{\mathrm{EAF},i,j,k}$ of all other regions with $(i,j,k)\neq (1,1,1)$, the nonlinear fits hold high uncertainties.
In particular, this affects all regions of the northern hemisphere, $A_\mathrm{NH}$.
Low solar irradiation around the perihelion passage results in low dynamical activity and weak signals for these locations. 
In addition, the discussion of uncertainties remains unchanged once the shape resolution of $\Delta x\approx 150\,\mathrm{m}$ is moderately decreased or increased.
We consider two types of TPM, model A and model $\alpha$.
The qualitative discussion does not differ for both model types and
for the values $\alpha=1.5$ and $\alpha=2$.
In general, the temporal sampling and the uncertainties prescribed for the $\omega$ data affect the fit.
Our discussion does not change for a data set with modified sampling and uncertainty weights.

The same control for the rotation state is possible for each combination between the two shapes, the bilobed shape and the shape 67P, and the two TPMs, model A and model $\alpha$.
This ability is a characteristic of the domain decomposition.
For lowest-energy states and sublimation following an instantaneous TPM, each region is related to torque formation pointing into one of the eight spatial quadrants with respect to the solar basis.
Each region collects one specific combination of (signs for) the vectorial torque efficiency.
Thus, the weights of the effective active fraction control all torque directions.
Eight is the minimum number for weights (and regions), and it is obtained by our approach.

On the surface of comet 67P/C-G for a wide range of locations, the weights for effective active fraction are not well constrained by rotation data.
This indetermination stems from the specifics of both, the geometry of the nucleus and the activity model.
We have shown this comparing results for two shapes and two activity models.
A better localization of the activity must consider complementary data.
\cite{Attree2023a} shows the option to add NGA to form a joint data set.
Globally integrated gas production such as in \cite{Lauter2020} and locally resolved gas activity in \cite{Lauter2022} can also be included.
Any approach for a joint data set must address the difficulty of a relative weighting of the different data sets, which introduces a bias to the final result.
With knowledge of the shape geometry, the localization of surface activity can be applied to other solar system objects that have weaker sources of torque activity.
Nongravitational momenta have this potential on dark comets reported by \cite{Seligman2023} as well as the YORP effect on asteroids reviewed by \cite{Bottke2015}.

\begin{acknowledgements}
The authors gratefully acknowledge the computing time made available
to them on the high-performance computer Lise at the NHR center NHR@ZIB.
This center is jointly supported by the federal ministry of education and research and the state governments participating in the NHR.
This research was supported by the International Space Science Institute (ISSI) in Bern, through ISSI International Team project \#547 (Understanding the Activity of Comets Through 67P’s Dynamics)
with N.~Attree acting as PI.
M.L. acknowledges the support by Johannes Kepler Universit{\"a}t Linz Austria for a research visit in 2024.
We thank the referee for the helpful comments, which improved the presentation of the manuscript.
\end{acknowledgements}

\begin{appendix}

\section{Geometric properties}

For the shape model of comet 67P given by \cite{Preusker2017}, \cite{Kramer2019a} have derived the tensor of inertia.
Based on a strictly homogeneous density distribution within the body their tensor suffers from an incompatibility of the local basis vector, $e_3$, and the observed angular velocity vector.
To circumvent this issue they add inhomogeneities to the density and obtain an adapted tensor of inertia,
\begin{equation}\label{eq:tensor}
\begin{pmatrix}
9.3408457 \times 10^{18} \hspace{2mm} 5.6695663 \times 10^{16}\qquad\qquad  0 \\
5.6695663 \times 10^{16} \hspace{2mm} 1.6562414 \times 10^{19}\qquad\qquad  0 \\
0\qquad\qquad\qquad\quad{}0\qquad\qquad 1.8192083 \times 10^{19}
\end{pmatrix}\,
\mathrm{kg}\, \mathrm{m}^2,
\end{equation}
which is applied to Sect.~\ref{sec:systems}.

At position $x\in\partial\Omega_t$, the torque density, $t_\mathrm{NG}$, in Eq.~\eqref{eq:ngt} has the direction of the term $x\times \nu$.
Two components with respect to the local basis, $B(\nu)$, in Fig.~\ref{fig:local} satisfy the relation
\begin{gather*}
(x\times \nu)\cdot b_1(\nu)
=
\frac{-\nu\cdot b_3}{\sqrt{1-(\nu\cdot b_3)^2}}
(x\times \nu)\cdot b_3.
\end{gather*}
This transfers to the signs of components for $t_\mathrm{NG}$.
\begin{equation}\label{eq:oneandthree}
\begin{aligned}
    \sgn{} ( t_\mathrm{NG}\cdot b_1(\nu) )
& = \sgn{} ( t_\mathrm{NG} \cdot b_3 )
~~ \text{on}~~ \partial\Omega_\mathrm{SH},
\\
\sgn{} (t_\mathrm{NG} \cdot b_1(\nu) )
& = - \sgn{} (t_\mathrm{NG} \cdot b_3 )
~~ \text{on}~~ \partial\Omega_\mathrm{NH}
\end{aligned}
\end{equation}

\section{Fourier modes for torque}\label{sec:fourier}

Fourier decompositions have been evaluated for NGA by \cite{Kramer2019} and for NGT by \cite{Kramer2019a}.
Both approaches apply Fourier modes for vector quantities, which then simplify in the rotating frame.
They share the same scalar field for mass generation.
For the present section we apply the Fourier decomposition to this scalar field only.
We consider specific assumptions for the rotation state.
First the solution, $R$ and $L$, of Eq.~\eqref{eq:rotation} describes a principal axis rotation around the short axis $b_3$ with constant $|\omega|$ and the rotation period $\Delta t = 2\pi/|\omega|$.
With elementary matrices $R_\mathrm{x}$ and $R_\mathrm{z}$ in \cite{Montenbruck2000} in this situation, $R$ can be represented with Euler angles by
\begin{gather*}
R(s) = R_\mathrm{z}(-\phi_0)R_\mathrm{x}(-\theta_0)R_\mathrm{z}(-\psi_0-|\omega| s)
\end{gather*}
at time $s$ with appropriate angles $\phi_0$, $\theta_0$, $\psi_0$.
In Eq.~\eqref{eq:ngt} torque, $T$, depends on the vector $r_\mathrm{s}$ pointing toward the sun.
The second assumption is to consider a constant $r_\mathrm{s}$.
This leads to a constant solar basis, $B_\mathrm{s} = B(r_\mathrm{s})$, in Eq.~\eqref{eq:sekanina}.
Because the orthogonal matrices $B_\mathrm{s}$ and $R(s)$ agree in the third column, we can find a time $t_0$ to bring into agreement $B_\mathrm{s} = R(t_0)$.
With $\lambda(s) = |\omega|(s-t_0)$, this leads to $B_\mathrm{s}^\mathrm{T}R(s) = R_\mathrm{z}(-\lambda(s))$.

At position $\xi\in\partial\Omega$ with the outward normal $\nu_\xi$ in the body frame, the vector field for torque in Eq.~\eqref{eq:ngt} can be expressed by $t_\mathrm{NG}(x(s),s) = B_\mathrm{s}\, t_\mathrm{BF}(B_\mathrm{s}^\mathrm{T}x(s))$ with $x(s) = R(s)\xi$ and the torque in the body frame,
\begin{gather*}
t_\mathrm{BF}(\xi) = - \dot m_\mathrm{BF}(\nu_\xi)\, \xi \times \nu_\xi
,\quad
\dot m_\mathrm{BF}(\nu_\xi) = v_\mathrm{gas}\,
\dot m(I(B_\mathrm{s}\nu_\xi,r_\mathrm{s})).
\end{gather*}
For the time average in Eq.~\eqref{eq:taverage}, we obtain
\begin{equation}\label{eq:torqueaverage}
\begin{aligned}
\overline{t_\mathrm{NG}}(x(t),t)
& = \frac{B_\mathrm{s}}{\Delta t} \intl_{t}^{t+\Delta t}
t_\mathrm{BF}(B_\mathrm{s}^\mathrm{T} x(s))\,ds
\\ & =
\frac{B_\mathrm{s}}{2\pi} \intl_{0}^{2\pi}
t_\mathrm{BF}(R_\mathrm{z}(-\lambda)\xi)\,d\lambda.
\end{aligned}
\end{equation}
The torque $t_\mathrm{BF}(R_\mathrm{z}(-\lambda)\xi)$ can be separated into a product of the scalar term $f(\lambda) = \dot m_\mathrm{BF}(R_\mathrm{z}(-\lambda)\nu_\xi)$ and the vectorial component $-R_\mathrm{z}(-\lambda)(\xi\times \nu_\xi)$.
During one rotation the angle of maximum momentum is related to the angle
\begin{gather*}
\lambda_\mathrm{M} = \argmax_{0 \leq \lambda \leq 2\pi} f(\lambda).
\end{gather*}
The value of $\lambda_\mathrm{M}$ is specific for $\xi$ and we denote the corresponding time $t_\mathrm{M}$ holding the property $\lambda(t_\mathrm{M}) = \lambda_\mathrm{M}$.
Because of monotonicity for $\dot m$ together with irradiation, $I$, in Eq.~\eqref{eq:idef}, $t_\mathrm{M}$ can be expressed by $\argmax_s(r_\mathrm{s} \cdot R(s)\nu_\xi)$, equivalently.
We observe two properties for this maximum.
For the local basis in Eq.~\eqref{eq:sekanina} the maximization property of $t_\mathrm{M}$ yields the relation $B(\nu(t_\mathrm{M})) = B_\mathrm{s}$, which gives
\begin{equation}\label{eq:crosstrafo}
R_\mathrm{z}(-\lambda_\mathrm{M})(\xi\times\nu_\xi) = B^\mathrm{T}(\nu(t))\,(x(t) \times \nu(t)).
\end{equation}
Secondly $f(\lambda)$ is a symmetric function with respect to $\lambda_\mathrm{M}$.
That is why function $f(\lambda)$ has vanishing coefficients related to $\sin$ functions and it has the Fourier coefficients
\begin{equation*}
c'_n(\xi,r_\mathrm{s}) =
\frac{1}{\pi} \intl_0^{2\pi}\cos(n(\lambda-\lambda_\mathrm{M}))
f(\lambda)\,d\lambda.
\end{equation*}
Following \cite{Chesley2004} for spherical geometry, irradiation, $I(B_\mathrm{s} R_\mathrm{z}(-\lambda)\nu_\xi,r_\mathrm{s})$, can be expressed representing the diurnal trend at the position $\xi$.
In general $f(\lambda)$ is nonlinear and it reflects the properties of the TPM in Sect.~\ref{sec:tpm}.
The coefficients $c'_n$ are the result of numerical integration resulting in
\begin{gather*}
f(\lambda) =
\frac{c'_0}{2} + \sum_{n=1}^\infty c'_n \cos(n(\lambda-\lambda_\mathrm{M})),
\\
\frac{1}{\pi}
\intl_0^{2\pi} f(\lambda) R_\mathrm{z}(-\lambda)\,d\lambda =
\diag(c'_1,c'_1,c'_0) \, R_\mathrm{z}(-\lambda_\mathrm{M}).
\end{gather*}
Together with Eq.~\eqref{eq:crosstrafo} this can be used to evaluate Eq.~\eqref{eq:torqueaverage}, which gives
\begin{equation}\label{eq:fouriertrafo}
b_{i}(r_\mathrm{s}) \cdot \overline{t_\mathrm{NG}}(x,t)
= - c_i \, b_i(\nu) \cdot (x \times \nu)
\end{equation}
with $c_1 = c_2 = c'_1/2$ and $c_3 = c'_0/2$.
As a consequence the torque average, $\overline{t_\mathrm{NG}}(x(t),t)$, is specific for the position $\xi$, and it is constant in time $t$.

For example at position $\xi\in\partial\Omega$, the specific choice
\begin{gather*}
f(\lambda) =
\begin{cases}  
\dot m_0 & \text{for}\quad
|\lambda-\lambda_M|\leq \pi/2
\\
0 & \text{otherwise}
\end{cases}  
\end{gather*}
represents a constant activity along the rotating position $x(t)$ with some constant $\dot m_0=\dot m_0(\xi)$.
This idealized assumption yields the result
$c_1 = c_2 = \dot m_0/\pi$, $c_3 = \dot m_0/2$.

\end{appendix}

\end{document}